\documentclass[11pt,a4paper]{article}

\pdfoutput=1

\usepackage{jheppubnohead}
\usepackage{epsfig}
\usepackage{latexsym}
\usepackage{hyperref}
\usepackage{graphics,graphicx,color,amssymb,float,amsmath}
\usepackage{subfig}
\usepackage{pstricks}
\usepackage{color}
\usepackage{mathrsfs}
\usepackage{xspace}
\usepackage{listings}
\usepackage{booktabs}
\usepackage{bbold}
\usepackage{slashed}
\usepackage{siunitx}

\usepackage{makecell, tabularx}

\setcellgapes{3pt}

\usepackage[capitalise, english]{cleveref}
\crefname{section}{Sec.}{Secs.}
\crefname{table}{Tab.}{Tabs.}
\def\be{\begin{equation}}
\def\ee{\end{equation}}

\newcommand{\keff}{\kappa_{\rm eff}}
\newcommand{\keffi}{\kappa_{\rm eff,i}}
\newcommand{\lam}{{\bf \lambda}}
\newcommand{\maxino}{m_{\widetilde a}}
\newcommand{\mixH}{x_{\widetilde a , \widetilde H}}
\newcommand{\mixB}{x_{\widetilde a , \widetilde B}}
\newcommand{\mixN}{x_{\widetilde a  \widetilde \nu}}
\newcommand{\tev}{{\rm TeV}}

\newcommand{\mev}{{\rm MeV}}
\newcommand{\kev}{{\rm keV}}
\subheader{BONN-TH-2018-10}

\title{The Supersymmetric R-Parity Violating Dine-Fischler-Srednicki-Zhitnitsky Axion Model}

\author[a]{Stefano Colucci,}
\author[a]{Herbi K. Dreiner,}
\author[b]{Lorenzo Ubaldi}

\affiliation[a]{Physikalisches Institut der Universit\"at Bonn, Bethe Center
for Theoretical Physics, \\ Nu{\ss}allee 12, 53115 Bonn, Germany}

\affiliation[b]{SISSA International School for Advanced Studies, \\
Via Bonomea 265, 34136 Trieste, Italy}

\emailAdd{colucci@th.physik.uni-bonn.de}
\emailAdd{dreiner@th.physik.uni-bonn.de}
\emailAdd{ubaldi.physics@gmail.com}

\abstract{
We propose a complete R-parity violating supersymmetric model with baryon triality that contains a Dine-Fischler-Srednicki-Zhitnitsky 
axion chiral superfield.  We parametrize supersymmetry breaking with soft terms, and determine under which conditions the model is 
cosmologically viable. As expected we always find a region of parameter space in which the axion is a cold dark matter candidate. The 
mass of the axino, the fermionic partner of the axion, is controlled by a Yukawa coupling. When heavy [${\cal O}$(TeV)], the axino decays 
early and poses no cosmological problems. When light [$\mathcal{O}$(keV)] it can be long lived and a warm dark matter candidate. We 
concentrate on the latter case and study in detail the decay modes of the axino. We find that constraints from astrophysical X- and 
gamma rays on the decay into photon and neutrino can set new bounds on some trilinear supersymmetric R-parity violating Yukawa
couplings. In some corners of the parameter space the decays of a relic axino can also explain a putative 3.5 keV line.
}

\begin{document}

\maketitle

\section{Introduction} \label{sec:Introduction}

Supersymmetry (SUSY) is the best candidate to explain the hierarchy problem of the Standard Model (SM) 
\cite{Gildener:1976ai,Veltman:1980mj}, and thus a leading candidate for the discovery of new physics effects at the LHC. 
Most supersymmetric models which have been searched for at the LHC make the assumption of conserved R-parity 
\cite{Farrar:1978xj}. A very restrictive and widely considered version with universal boundary conditions at the unification 
scale, the constrained minimal supersymmetric Standard Model (CMSSM) \cite{Kane:1993td}, is now in considerable tension with 
collider data \cite{Bechtle:2014yna}. The R-parity violating (RpV) CMSSM is still very much allowed \cite{Dercks:2017lfq}.

R-parity is a discrete multiplicative symmetry invoked in the CMSSM  to forbid baryon and lepton number violating operators 
which together lead to rapid proton decay. The bonus of imposing 
R-parity is that the lightest supersymmetric particle (LSP), typically the neutralino, is stable and provides a dark matter 
(DM) candidate in the form of a weakly interacting massive particle (WIMP). This is the most popular and most searched 
for DM candidate. However, to--date no neutralino dark matter has been found \cite{Aprile:2017iyp, Akerib:2016vxi, Tan:2016zwf}. 
It is thus prudent to investigate other DM candidates, also within alternative 
supersymmetric models.

R-parity does not forbid some dimension-5 operators dangerous for proton decay~\cite{Weinberg:1981wj}. This issue 
can be resolved by imposing a $Z_6$ discrete symmetry, known as proton hexality (P$_6$), which leads to the 
same renormalizable superpotential as R-parity~\cite{Dreiner:2005rd,Dreiner:2007vp}, but is incompatible with a 
GUT symmetry. Alternatively, one can impose a $Z_3$--symmetry known as baryon-triality (B$_3$) 
\cite{Dreiner:2005rd,Ibanez:1991hv,Martin:1997ns,Grossman:1998py,Dreiner:2006xw,Dreiner:2007uj,Bernhardt:2008jz}. 
The latter allows for lepton number violating operators in the superpotential thanks to which the neutrinos acquire Majorana 
masses, without introducing a new very high energy Majorana mass scale~\cite{Allanach:2003eb,Grossman:1998py,Dreiner:2006xw,Dreiner:2007uj,Hall:1983id,Nardi:1996iy,Abada:1999ai,Hirsch:2000ef,Davidson:2000uc}. 
This is a virtue of B$_3$ models.\!\footnote{Other than P$_6$ and B$_3$, one can also consider $R$-symmetries to restrict 
the renormalizable Lagrangian resulting in R-parity conservation or violation 
\cite{Chamseddine:1995gb,Lee:2010gv,Lee:2011dya,Dreiner:2013ala,Chen:2014gua}. For our purposes this is equivalent.}  
A possible handicap is that the LSP is unstable and is not a dark matter candidate. This can be naturally resolved when 
taking into account the strong CP problem, as we discuss in the following.  

Besides the hierarchy problem, every complete model should also address the  strong CP problem~\cite{Peccei:2006as}, 
which plagues the SM, as well as its supersymmetric generalizations. In its simplest forms this necessitates 
introducing the pseudo scalar axion field~\cite{Wilczek:1977pj, Weinberg:1977ma}. In the supersymmetric versions, the 
axion is part of a chiral supermultiplet and is accompanied by another scalar, the saxion, and a spin-1/2 fermion, the axino. 

In this paper we propose to study a B$_3$ model with the inclusion of an axion supermultiplet of the 
Dine-Fischler-Srednicki-Zhitnitsky (DFSZ) type~\cite{Dine:1981rt, Zhitnitsky:1980tq}. The axion contributes to the DM energy 
density in the form of cold DM \cite{Preskill:1982cy,Holman:1982tb}. Depending on the value of its decay constant, $f_a$, it 
constitutes all or a fraction of the DM. The gravitino is heavy [${\cal O} ({\rm TeV})$], decays early in the history of the universe, 
and does not pose cosmological problems.  The axino mass is proportional to $\lambda_\chi v_\chi$ [see 
Eq.~\eqref{eq:axinomass}], where $v_\chi$ is the vacuum expectation value (VEV) of a scalar field of the order of the soft 
masses [${\cal O}$(TeV)], and $\lambda_\chi$ is a dimensionless Yukawa coupling in the superpotential. We consider two cases: 
(i) a heavy axino ($\sim$ TeV, which requires $\lambda_\chi \sim 1$), and (ii) a light one ($\sim$ keV, requiring $\lam
_\chi \ll 1$). In case (i) the axino decays early and is not a DM candidate. In the more interesting case~(ii), it is the LSP, its 
lifetime is longer than the age of the universe, and contributes as warm DM \cite{Rajagopal:1990yx}. We study in detail its three
decay channels: into an axion and a neutrino, into three neutrinos, into a neutrino and a photon, the last one subject to 
constraints from X- and gamma-ray data.

In 2014 a potential anomaly was observed in X-ray data coming from various galaxy clusters and the Andromeda galaxy 
\cite{Bulbul:2014sua, Boyarsky:2014jta}. There have been several papers discussing this in terms of an axino
in R-parity violating supersymmetry \cite{Kong:2014gea,Choi:2014tva,Liew:2014gia}. Although we were able to show that
this explanation is not probable \cite{Colucci:2015rsa}.

The outline of the paper is as follows. In Sec.~\ref{sec:model}, 
we introduce the model, describe its parameters and mass 
spectrum. In Sec.~\ref{sec:axino-decays} we compute in detail the axino decay rates and branching fractions. In 
Sec.~\ref{sec:cosmology} we consider cosmological and astrophysical constraints on the model, with a focus on the more 
interesting case of a light axino. We conclude with a discussion of our results and point out possible future directions of investigation.

In Appendix~\ref{sec:lightaxino} we include some general comments on the axino mass, we discuss its dependence 
on the SUSY breaking scale and point out that DFSZ models accommodate more easily a light axino, as opposed
to models \emph{\`a la} Kim-Shifman-Vainshtein-Zakharov (KSVZ)~\cite{Kim:1979if, Shifman:1979if}. In 
Appendix~\ref{sec:axino-mixing} we furthermore
give details on the derivation of the mixings of the axino mass eigenstates.

\section{The RpV DFSZ Supersymmetric Axion Model } \label{sec:model}
\subsection{The Lagrangian}
\begin{table}[t]
\centering
  \begin{tabular}{ccccc}\toprule
    Superfield & SU(3)$_C$ & SU(2)$_L$ & U(1)$_Y$ & ~U(1)$_{\text PQ}~$  \\ \midrule
    $\hat Q$ & $3$ & $2$ & $1/6$ & $Q_{Q}$  \\ 
    $\hat{\bar U}$ & $\bar 3$ & $ 1$ & $-2/3$ & $Q_{\bar U}$  \\ 
    $\hat{\bar D}$ & $\bar 3$ & $ 1$ & $1/3$ & $Q_{\bar D}$  \\ 
    $\hat L$ & $1$ & $2$ & $-1/2$ & $Q_{L}$  \\ 
    $\hat{\bar E}$ & $1$ & $1$ & $1$ & $Q_{\bar E}$  \\ 
    $\hat H_d$ & $1$ & $2$ & $-1/2$ & $Q_{H_d}$  \\ 
    $\hat H_u$ & $1$ & $2$ & $1/2$ & $Q_{H_u}$ \\ 
    $\hat A$ & $1$ & $1$ & $0$ & $Q_{A}$  \\  
    $\hat {\bar A}$ & $1$ & $1$ & $0$ & $-Q_{A}$  \\ 
    $\hat \chi$ & $1$ & $1$ & $0$ & $0$  \\ \bottomrule
  \end{tabular}
   \caption{Charge assignments for the chiral
    superfields. Note in our R-parity violating supersymmetric models $Q_{H_d}=Q_L$.}
  \label{tab:charges}
\end{table}
The building blocks of the minimal supersymmetric DSFZ axion model are the particle content of the MSSM plus three superfields, 
which are necessary to construct a self-consistent axion sector as discussed in Ref.~\cite{Dreiner:2014eda}. The particle content, 
the quantum numbers with respect to the SM gauge sector $SU(3)_c \times SU(2)_L \times U(1)_Y$, and the charges under the 
global Peccei-Quinn (PQ) symmetry~\cite{Peccei:1977hh}, $U(1)_{\rm PQ}$, are summarized in Table~\ref{tab:charges}. The 
renormalizable superpotential is given by
\begin{align}
W & =  W_{B_3}  + W_I + W_{\rm PQ} \, , \label{eq:superpotential} \\
W_{B_3} & = Y_u \hat Q \hat H_u \hat {\bar U} + Y_d \hat Q \hat H_d \hat {\bar D} + Y_e \hat L \hat H_d \hat {\bar E}
 + \frac{1}{2} \lambda \hat L \hat L \hat {\bar E} + \lambda^{'} \hat L \hat Q \hat {\bar D}  \, , \label{eq:WB3} \\
 W_I & = c_1 \hat A \hat H_u \hat H_d + c_2 \hat A \hat L \hat H_u \, , \label{eq:WI} \\
 W_{\rm PQ} & =  \lambda_\chi \hat \chi \left( \hat A \hat {\bar A} - \frac{1}{4} f_a^2 \right) 
  \label{eq:WPQ} \, .
\end{align}
Here and in the following we suppress generation, as well as isospin and color indices. Superfields are denoted by a hat 
superscript. We have imposed the discrete $Z_3$ symmetry~\cite{Dreiner:2005rd} known as baryon triality, $B_3$. This leaves 
only the RpV operators $\frac{1}{2}\lam \hat 
L \hat L \hat {\bar E}$, $\lambda^{'} \hat L \hat Q \hat {\bar D}$, and $c_2 \hat A \hat L \hat H_u$. We have generalized the usual
bilinear  operators $\mu H_u H_d$ and $\kappa_i L_i H_u$ to obtain PQ invariance. The PQ charges satisfy
\be
Q_A + Q_{H_u} + Q_{H_d} = 0 \, , \quad Q_A + Q_{H_u} + Q_{L} = 0 \, ,
\ee
from the terms in Eq.~\eqref{eq:WI}, and similar relations for the other fields that are readily obtained from 
the terms in the rest of the superpotential.

The PQ symmetry  is spontaneously broken at the scale $f_a$, due to scalar potential resulting from $W_{\rm PQ}$,
and the scalar parts of $\hat A$ and $\hat {\bar A}$ get vacuum expectation values (VEVs) $v_A,\,v_{\bar A}$, respectively,
\begin{equation}
A = \frac{1}{\sqrt{2}}\left(\phi_A + i \sigma_A + v_A \right)\,,\hspace{1cm} \bar{A} = \frac{1}{\sqrt{2}}\left(\phi_{\bar A} + i \sigma_{\bar A} + v_{\bar A} \right)\, .
\label{eq:axion-fields}
\end{equation}
The VEVs must fulfill 
\begin{equation}
v_A \cdot v_{\bar A} = \frac12 f_a^2  \, ,
\end{equation}
and we denote their ratio as
\begin{equation}
 \tan^2 \beta' \equiv \frac{v_{\bar A}}{v_A} \, .
\end{equation}
Below the PQ breaking scale, we generate an effective $\mu$-term, $\mu_{\rm eff} \hat H_u \hat H_d$, and effective bilinear RpV $\kappa$-terms, $\kappa_{\rm eff,i} \hat L_i \hat H_u$, with 
\begin{equation}
\mu_{\rm eff} = \frac{c_1}{\sqrt{2}} v_A \, , \hspace{1cm} 
\keffi = \frac{c_{2,i}}{\sqrt{2}} v_A \, .
\end{equation}
For successful spontaneous breaking of the electroweak (EW) symmetry, $\mu_{\rm eff}$ should be of the order of the EW scale, while 
$\kappa_{\rm eff}$ is constrained by neutrino physics, as discussed below, and should be $\leq O(\text{MeV})$. Since 
$f_a > 10^9$ GeV from astrophysical bounds on axions~\cite{Raffelt:2006cw}, the couplings $c_1$ and $c_2$ must be very 
small, roughly $c_1 < 10^{-6}$ and $c_2 < 10^{-11}$. They are, nonetheless, radiatively stable.

We have assumed that $W_{\rm PQ}$ respects an $R$-symmetry, under which the field $\hat \chi$ has charge 2. This 
forbids the superpotential terms $\hat \chi^2$ and $\hat \chi^3$. See also Ref.~\cite{Chamseddine:1995gb}.

The soft-supersymmetric terms of this model consist of scalar squared masses, gaugino masses, and the counterparts 
of the superpotential couplings. The full soft Lagrangian reads
\begin{align}
- \mathcal{L}_{\rm soft} = & (M_1  \widetilde B \widetilde B + M_2 \widetilde W \widetilde W + M_3 \widetilde g \widetilde g 
+ {\rm h.c.}) +  \widetilde{f}^ \dagger m^2_{\widetilde{f}} \widetilde{f} \nonumber  \\
 &+  m^2_{\chi} |\chi|^2 + m^2_{{A}} |A|^2 + m^2_{\bar{{A}}} |\bar{A}|^2 + m^2_{H_u} |H_u|^2 + m^2_{H_d} |H_d|^2 
 + m^2_{\ell H_u}(\widetilde{\ell}^{\dagger}H_u + H_u^{\dagger} \widetilde{\ell}) \nonumber  \\
 &+\left(B\mu H_d \cdot H_u + \rm h.c.\right)
 +  (T_u \widetilde{u} \widetilde{q} H_u + T_d \widetilde{d} \widetilde{q} H_d + T_e \widetilde{e} \widetilde{\ell} H_d + 
 T_{\lambda} \widetilde\ell \widetilde \ell \widetilde e + T_{\lambda^{'}} \widetilde\ell \widetilde q \widetilde d  \nonumber \\
&  + T_{c1} A H_u H_d + T_{c2} A \,\widetilde{\ell} H_u +  T_{\lambda_\chi} \,\chi A \bar{A} + L_V \chi  + {\rm h.c.})  \;,
\label{softLag}
\end{align}
with $\widetilde{f}\in \{\widetilde{e}, \widetilde{\ell},\widetilde{d},\widetilde{u},\widetilde{q} \}$. 
We assume that SUSY breaking violates the $R$-symmetry of the axion sector, hence we include the soft terms $T_{\lambda_\chi} 
\,\chi A \bar{A}$ and $L_V \chi$. Note that, even if set to zero at some scale, these terms would be generated at the two-loop level 
because of the small coupling to the MSSM sector, where no $R$-symmetry is present. 

At low energy the Lagrangian contains interactions of the axion with the gauge fields. These are best understood by performing a field 
rotation\footnote{See, for example, Appendix C of Ref.~\cite{DEramo:2015iqd} for details.} that leads to a basis in which all the matter 
fields are invariant under a PQ transformation~\cite{Georgi:1986df}. We have
\begin{equation}
\label{eq:LaVV}
\mathcal{L}_{aVV} = \frac{a}{32 \pi^2 f_a}   \left(g_1^2 C_{aBB}  B_{\mu\nu} \widetilde B^{\mu\nu} +  g_2^2 C_{aWW} W^a_{\mu\nu} 
\widetilde W^{a \mu\nu} + g_3^2  G^\alpha_{\mu\nu} \widetilde G^{\alpha \mu\nu} \right)\, .
\end{equation}
Here $a$ is the axion field which in our model is given by $a \simeq \frac{1}{\sqrt{2}} (\sigma_A - \sigma_{\bar A})$ to a good 
approximation\footnote{The axion also has a small admixture of $H_u$ and $H_d$ which we neglect here.}\!\!, \textit{cf.} 
Eq.~(\ref{eq:axion-fields}). $B_{\mu\nu}$, $W^a_{\mu\nu}$, $G^a_{\mu\nu}$ are the SM field strength tensors, here $a$ 
is an adjoint gauge group index, with $\widetilde X^{a\mu\nu} \equiv \epsilon^{\mu\nu\rho\sigma} X^a_{\rho\sigma}$, and 
the coefficients are given in terms of the PQ charges
\begin{align}
C_{aBB} & = 3(3Q_Q + Q_L) - 7 Q_A  \, , \\
C_{aWW} & = - 3(3Q_Q + Q_L) + Q_A  \, .
\end{align}
In the supersymmetric limit the couplings of the axino to gauginos and gauge fields are related to those in Eq.~\eqref{eq:LaVV}. We 
are, however, interested in the case of broken SUSY. In general, such axino couplings can be calculated explicitly, by computing 
triangle loop diagrams, once the full Lagrangian is specified, as it is in our model.  Later in the paper we use the full Lagrangian to 
compute the one-loop decay of the axino into a photon and a neutrino, including the couplings.

\subsection{The Mass Spectrum}
We now turn our attention to the spin-1/2 neutral fermion mass spectrum of this model. The $10\times10$  mass 
matrix in the 
basis  $(\lambda_{\widetilde B}, \widetilde W^0, \widetilde H_u^0, \widetilde H_d^0, \nu_{L,i}, \widetilde A, \widetilde {\bar A}, \widetilde 
\chi)$, which is a generalization of the MSSM neutralino mass matrix, reads
 \begin{equation} \label{eq:Mchi10}
\mathcal{M}_{\chi^0} = \left[ 
\begin{array}{cccccccccc}
M_1 &0 &\frac{g_1 v_u}{2}   &-\frac{g_1 v_d}{2}  &0 &0 &0 &0 &0 &0\\
0 &M_2 &-\frac{g_2 v_u}{2}   &\frac{g_2 v_d}{2}  &0 &0 &0 &0 &0 &0\\
\frac{g_1 v_u}{2}   &-\frac{g_2 v_u}{2}  &0 &- \frac{c_1 v_A}{\sqrt{2}}  &\frac{c_{2,1} v_A}{\sqrt{2}}  &\frac{c_{2,2} v_A}{\sqrt{2}}  
&\frac{c_{2,3} v_A}{\sqrt{2}}  &- \frac{c_1 v_d}{\sqrt{2}}  &0 &0\\
-\frac{g_1 v_d}{2}   &\frac{g_2 v_d}{2}   &- \frac{c_1 v_A}{\sqrt{2}}   &0 &0 &0 &0 &- \frac{c_1 v_u}{\sqrt{2}}  &0 &0 \\
0 &0 &\frac{c_{2,1} v_A}{\sqrt{2}}   &0 &0 &0 &0 &\frac{c_{2,1} v_u}{\sqrt{2}}  &0 &0\\
0 &0 &\frac{c_{2,2} v_A}{\sqrt{2}}   &0 &0 &0 &0 &\frac{c_{2,2} v_u}{\sqrt{2}}  &0 &0\\
0 &0 &\frac{c_{2,3} v_A}{\sqrt{2}}   &0 &0 &0 &0 &\frac{c_{2,3} v_u}{\sqrt{2}}  &0 &0\\
0& 0&- \frac{c_1 v_d}{\sqrt{2}}   &- \frac{c_1 v_u}{\sqrt{2}}  & \frac{c_{2,1} v_u}{\sqrt{2}} & \frac{c_{2,2} v_u}{\sqrt{2}} & 
\frac{c_{2,3} v_u}{\sqrt{2}}  &0 &\frac{v_{\chi} \lambda_\chi}{\sqrt{2}}   &\frac{v_{\bar{A}} \lambda_\chi}{\sqrt{2}} \\
0& 0& 0& 0& 0& 0& 0& \frac{v_{\chi} \lambda_\chi}{\sqrt{2}}  &0 &\frac{v_A \lam_\chi}{\sqrt{2}}  \\
0& 0& 0& 0& 0& 0& 0& \frac{v_{\bar{A}} \lambda_\chi}{\sqrt{2}}  &\frac{v_A \lam_\chi}{\sqrt{2}}  &0 \\
\end{array} 
\right] \,.
\end{equation} 
$\mathcal{M}_{\chi^0}$ was obtained by first implementing our model with the computer program 
\texttt{SARAH}~\cite{Staub:2008uz,Staub:2009bi,Staub:2010jh}, then setting the sneutrino VEVs to zero, which 
corresponds to choosing a specific basis~\cite{Davidson:1996cc,Dreiner:2003hw}.
This is possible, since the superfields $\hat H_d$ and $\hat L_i$ have the same quantum numbers in RpV models. The choice of basis is 
scale dependent, in the sense that sneutrino VEVs are generated by renormalization group (RG) running, even if set to zero at a given 
scale. Hence, one needs to specify at what scale the basis is chosen. For later convenience, when we compute the axino decays, we 
choose this scale to be the axino mass. Neglecting for a moment the axino sector, which couples very weakly to the rest, and Takagi 
diagonalizing the upper-left $7\times 7$ block \cite{Dreiner:2008tw}, one finds~\cite{Hall:1983id,Grossman:1999hc} two massless 
neutrinos and a massive one with
\be
m_\nu \approx m_Z \cos^2 \beta \sin^2\xi \approx m_Z \cos^2\beta \ \frac{\kappa_{\rm eff}^2}{\mu_{\rm eff}^2} \, ,
\ee 
where
\be
\tan\beta = \frac{v_u}{v_d} \, , \quad \mathrm{and} \quad \cos\xi = \frac{\mu_{\rm eff}}{\sqrt{\mu_{\rm eff}^2+ \kappa_{\rm eff,1}^2 
+ \kappa_{\rm eff,2}^2
+ \kappa_{\rm eff,3}^2}} \, .
\ee
Requiring $m_\nu \approx 0.1$ eV, with $\tan\beta \approx 1$ and $\mu_{\rm eff} \approx 1$ TeV, one finds that $\kappa_{\rm eff,i}$ 
is of order MeV, and correspondingly smaller for larger $\tan\beta$. The two neutrinos which are massless at tree level acquire
a small mass at one loop~\cite{Hirsch:2000ef,Davidson:2000uc,Allanach:2003eb}, which we have not included in the 10x10 
mass matrix above.

In addition to the neutrinos, there are four eigenstates, mainly built from the MSSM neutralinos $(\lambda_{\widetilde B}, \widetilde 
W^0, \widetilde H_u^0, \widetilde H_d^0)$, with masses between 100 and 1000 GeV, and three axino eigenstates with masses that 
can be calculated analytically in the limit $v_{A} = v_{\bar{A}} = \frac{f_a}{\sqrt{2}}$, corresponding to $\tan\beta'=1$, and 
$c_{1,2} \to 0$. One finds for the axinos \cite{Dreiner:2014eda}
\begin{equation} \label{eq:max}
- \frac{1}{\sqrt{2}} \lambda_\chi v_{\chi} \,, \ \  \frac{1}{2\sqrt{2}} \Big(\lambda_\chi v_{\chi} 
\pm \lambda_\chi \sqrt{v_\chi^2+4 f_a^2} 
\Big) \, .
\end{equation}
The lightest of these three states corresponds to the fermionic component of the linear combination of superfields $\frac{1}
{\sqrt{2}}(\bar{A}-A)$. It is interpreted as the axino with a mass  \cite{Dreiner:2014eda}\footnote{Negative fermionic masses 
can be eliminated by a chiral rotation.}
\begin{equation} \label{eq:axinomass}
m_{\widetilde a} \simeq -\frac{1}{\sqrt 2} \lambda_\chi v_\chi \, .
\end{equation}
For the more general case of $\tan\beta' \neq 1$, we can not give analytical expressions. Instead, we find at the lowest order in 
$v_\chi/f_a$ 
\begin{equation} \label{eq:axinomassapprox}
m_{\widetilde a} \simeq -\frac{\sqrt 2 \tan^2\beta'}{1+\tan^4\beta'} \lambda_\chi v_\chi + {\cal O}\left(\lambda_\chi v_\chi 
\frac{v_\chi^2}{f_a^2} \right) \, .
\end{equation}
We also give the power of the next term in the expansion, where $f_a$ enters. This is the main contribution to the axino mass, we 
neglect small corrections proportional to $c_1$ and $c_2$.
We expect $v_\chi$ to be at the soft SUSY breaking scale, ${\cal O}
$(TeV). The Yukawa coupling $\lam_\chi$ then is our main parameter to control $m_{\widetilde a}$. $\lambda_\chi$ is radiatively 
stable: if we set it small at tree level, it remains small. See also Ref.~\cite{Allanach:1999mh} for a quantitative treatment of 
Yukawa RGEs inn supersymmetric models. The most interesting case we consider later is that of a light axino, with $m_{\widetilde 
a} = {\cal O}$(keV), which means $\lam_\chi$ must be very small.

In order to determine the interactions of the axino, in particular the potential decay modes, we would like to estimate the mixing
between the axino and the neutrinos. To do this one first diagonalizes the upper 7x7 block, neglecting the axinos which are very weakly 
coupled. The massive neutrino eigenstate will be mostly a combination of the gauge eigenstates $\nu_{L,i}$, with small components of bino, 
wino and higgsinos. Next include the lower 3x3 axino block. The off-diagonal entries between the neutrino and axino blocks are $c_2 v_u
/\sqrt{2} \simeq \kappa_{\rm eff} v_u / f_a$. Now rotate the 3x3 axino block to the mass eigenstates. The axino is a linear combination of 
$A$ and $\bar A$, the other two heavier fermions are a combination of $A$, $\bar A$ and $\chi$. The important point is that the entries
in the 3x3 Takagi diagonalization axino matrix are of order one. This implies that the pieces in the off diagonal blocks remain
of order $\kappa_{\rm eff} v_u / f_a$ after rotating the axino block, and we obtain the form of the matrix
\be \label{axnu2matr}
\mathcal{M}_{\widetilde{a},\nu} = \begin{bmatrix} m_{\nu} & {\cal O} \left(\frac{\keff v}{f_a} \right) \\ {\cal O} \left(\frac{\keff v}{f_a} 
\right) & \maxino \end{bmatrix}\, .
\ee
From this we can read off the axino-neutrino mixing
\be \label{eq:a-n_mix}
x_{\widetilde a,\nu} \approx \frac{\keff v}{\maxino f_a} \, .
\ee
For the purpose of this estimate and those in Sec.~\ref{sec:cosmology}, we take $v\equiv\sqrt{v_u^2+v_d^2}\sim v_u 
\sim v_d$ to be the weak scale.

The axino-higgsino mixing can be estimated by observing that the higgsino mass is of order $\mu_{\rm eff}$, and the
off-diagonal element between axino and higgsino is of order $c_1 v = \frac{\mu_{\rm eff}}{f_a} v$. Thus, the mixing is
\be  \label{ahimix}
x_{\widetilde a,\widetilde H} \approx \frac{c_1 v}{\mu_{\rm eff}} = \frac{v}{f_a} \, .
\ee
Considering $M_1 \sim \mu_{\rm eff} \sim M_{\rm SUSY} = {\cal O}$(TeV), we can estimate the bino-higgsino mixing as
$g_1 \frac{v}{M_{\rm SUSY}}$. Multiplying this times $x_{\widetilde a,\widetilde H}$ we obtain the axino-bino mixing 
\be \label{abimix}
x_{\widetilde a,\widetilde B}\approx g_1 \frac{v^2}{M_{\rm SUSY} f_a}\,.
\ee

Next we briefly consider the scalar sector. It is instructive to start by considering the limit $c_{1,2} \rightarrow 0$, in which 
there is no mixing between the MSSM sector and the axion sector, and the axion block of the scalar squared mass matrix 
can easily be diagonalized. We find a massless axion and a saxion with a mass squared $\sim m_A^2$ or 
$m_{\bar A}^2$, the parameters in the soft Lagrangian [$m_A,\,m_{\bar A} = {\cal O}$(TeV), \textit{cf.} Eq.~(\ref{softLag})]. 
They are, respectively, the CP-odd and the CP-even mass eigenstates, corresponding to the combination $\frac{1}{\sqrt{2}}
(\bar{A}-A)$. The result is not appreciably altered once we turn on the small couplings $c_1$ and $c_2$. The other four real 
scalar degrees of freedom in this sector are heavy, with masses of order $\lambda_\chi f_a$. The masses of the scalars in 
the MSSM sector are to a very good approximation those already well studied in the RpV SUSY literature 
\cite{Allanach:2003eb,Allanach:1999mh}.

\section{Light Axino Decay Modes} \label{sec:axino-decays}
%
In this section we consider an axino mass lower than twice the electron 
mass, in which case the axino has only three three decay modes:
\begin{enumerate}
\item into a neutrino and an axion (via a dimension 5 operator), $\tilde a \to \nu_i + a$;
\item into three neutrinos (tree-level), $\tilde a \to \nu_i + \nu_j + \nu_k$;
\item into a neutrino and a photon (one-loop), $\tilde a \to \nu_i +\gamma$.
\end{enumerate}
We discuss them in turn.

\subsection{Decay into Neutrino and Axion}
\begin{figure}[t]
\centering
\subfloat[][]{{\label{fig:AxinoToAxionNu}\includegraphics[width=0.3\textwidth]{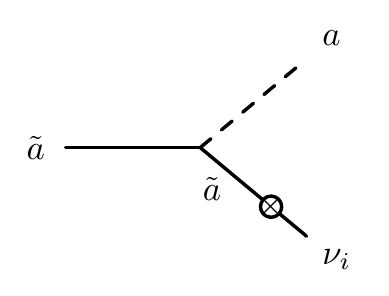}}}
\hspace{3cm}
\subfloat[][]{{\label{fig:AxinoTo3Nu}\includegraphics[width=0.3\textwidth]{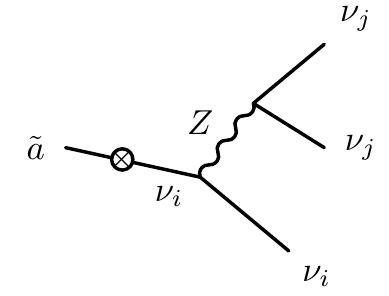}}}\\
\caption{\textit{Left}: Feynman diagram for the axino decay into an axion and a neutrino, as described by the five-dimensional 
operator of Eq.~\eqref{eq:dim5lag}. \textit{Right}: One of the Feynman diagrams describing the decay of the axino into three 
neutrinos. In our convention the $\otimes$  on a fermion line stands for a mixing between fermionic eigenstates. } 
\label{fig:TreeLevelDecays}
\end{figure}

The dominant operator responsible for the decay $\tilde a\to a+\nu$ is dimension 5:
\be \label{eq:dim5lag}
\frac{1}{f_a} (\partial^\mu a)  \bar \psi_{\tilde a} \gamma_\mu \gamma_5 \psi_{\tilde a} \, .
\ee
This is in the basis of the mass matrix in Eq.~(\ref{axnu2matr}), before we diagonalize to the final
mass eigenstate. Here $a$ is the axion, $\psi_{\tilde a}$ is the four-component spinor denoting the axino 
mass eigenstate and the neutrino arises due to mixing with the axino, once we diagonalize. A 
simple way to understand the origin of this operator is to consider the kinetic term $\bar \psi_{\tilde a} \gamma
_\mu \partial^\mu \psi_{\tilde a}$. After the chiral rotation $\psi_{\tilde a} \to e^{i \gamma_5 a / f_a} \psi_{\tilde a}$,
we obtain the above operator, when we expand the exponential. The diagram for the decay is shown in 
Fig.~\ref{fig:AxinoToAxionNu}, where we include the axino-neutrino mixing [see Eq.~\eqref{eq:a-n_mix}] in the final state.
The partial decay width into this channel is
\begin{align} \label{axnuwidth}
\Gamma_{\tilde a \to \nu a} & = \frac{x_{\tilde a ,\nu}^2}{16\pi} \frac{\maxino^3}{f_a^2} \approx \frac{1}{16\pi} \frac{v_u^2 \kappa_{\rm eff}^2}{f_a^4} \maxino \\
& = 6 \cdot 10^{-54} \ {\rm GeV} \left( \frac{\kappa_{\rm eff}}{\rm MeV} \right)^2 \left( \frac{10^{11} \ {\rm GeV}}{f_a}  \right)^4 \left( \frac{\maxino}{\rm keV}  \right) \, .
\end{align}
%
\subsection{Decay into Three Neutrinos}
This tree-level decay proceeds via the axino mixing with the neutrino and the exchange of a Z boson between the fermionic 
currents, as shown in Fig.~\ref{fig:AxinoTo3Nu}. Assuming for simplicity that the neutrino admixture of the axino is 
measured by $x_{{\widetilde a} ,\nu}$ for \textit{all} neutrino flavors, the decay width into all possible flavors of final neutrinos 
reads~\cite{Barger:1995ty}
\begin{align}
\Gamma_{\tilde a \to 3\nu} & = \frac{9 \alpha G_F^2}{8 \pi^4} x_{\tilde a ,\nu}^2 \maxino^5 \\
& = 3 \cdot 10^{-56} \ {\rm GeV} \  \left( \frac{\kappa_{\rm eff}}{\rm MeV} \right)^2 \left( \frac{10^{11} \ {\rm GeV}}{f_a}  
\right)^2 \left( \frac{\maxino}{\rm keV}  \right)^3 \, .
\end{align}
$\Gamma_{\tilde a \to 3\nu}$ and $\Gamma_{\tilde a \to \nu a}$ give the main contribution to the total decay width in most
of the parameter space. We find that  for $10^{10} \ {\rm GeV} < f_a < 10^{12} \ {\rm GeV}$ and $1 \ {\rm keV} < 
\maxino < 100 \ {\rm keV}$, the axino lifetime ranges from $10^{23}$ to $10^{41}$ s, which is longer than the
age of the universe ($\sim 4 \cdot 10^{17}$ s).

\subsection{Decay into Photon Plus Neutrino} \label{sec:2Bdecay}
\begin{figure}[t]
\centering
\subfloat[]{\raisebox{5ex}{\label{fig:LoopDiagFerW}\includegraphics[width=0.45\textwidth]{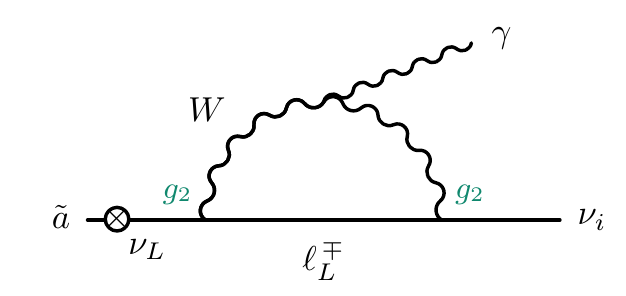}}}
\hspace{1cm}
\subfloat[][]{{\label{fig:LoopDiagBosW}\includegraphics[width=0.45\textwidth]{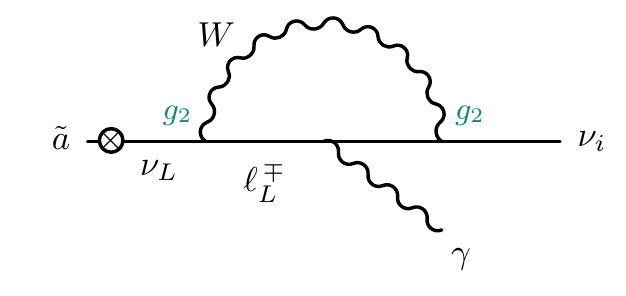}}}\\
\subfloat[]{\raisebox{5ex}{\label{fig:LoopDiagBosC}\includegraphics[width=0.45\textwidth]{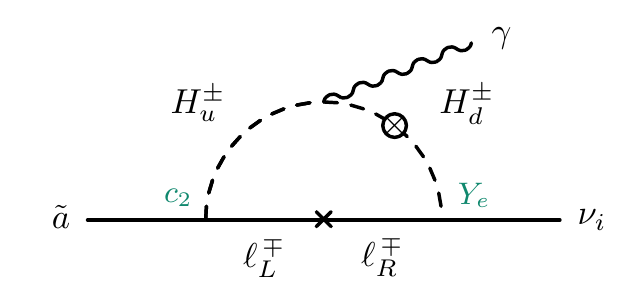}}}
\hspace{1cm}
\subfloat[][]{{\label{fig:LoopDiagFerC}\includegraphics[width=0.45\textwidth]{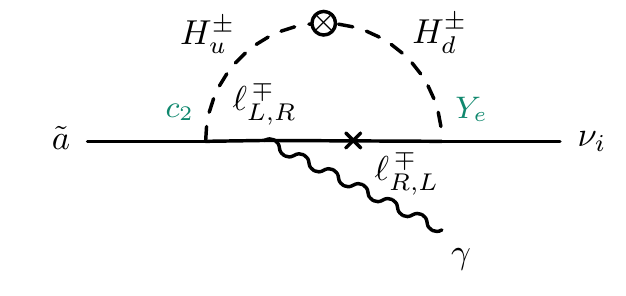}}}\\
\subfloat[]{\raisebox{5ex}{\label{fig:LoopDiagBos2C}\includegraphics[width=0.45\textwidth]{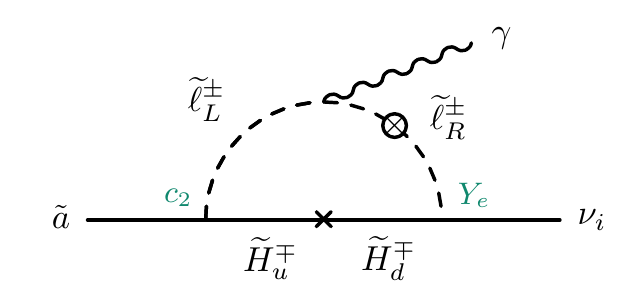}}}
\hspace{1cm}
\subfloat[][]{{\label{fig:LoopDiagFer2C}\includegraphics[width=0.45\textwidth]{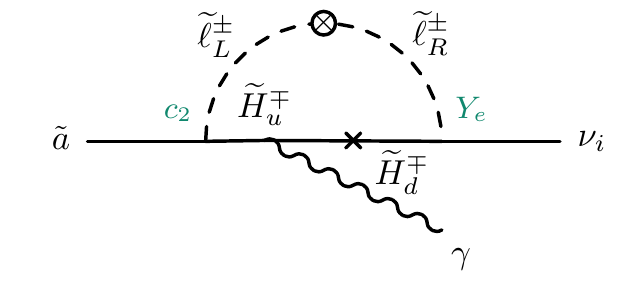}}}
\caption{Diagrams with no dependence on the the RpV trilinear couplings $\lambda^{(\prime)}$. The cross on a fermion 
line indicates a mass insertion needed for the chirality flip, while the $\otimes$ on a scalar or fermion line is a (dimensionless) mixing. We 
write the coupling constants in green at the vertices.} \label{fig:Diagrams_noRpV}
\end{figure}
\begin{figure}[h]
\centering
\subfloat[]{\raisebox{5ex}{\label{fig:LoopDiagBosB}\includegraphics[width=0.45\textwidth]{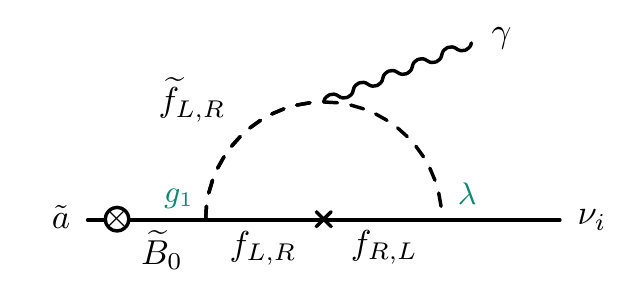}}}
\hspace{1cm}
\subfloat[][]{{\label{fig:LoopDiagFerB}\includegraphics[width=0.45\textwidth]{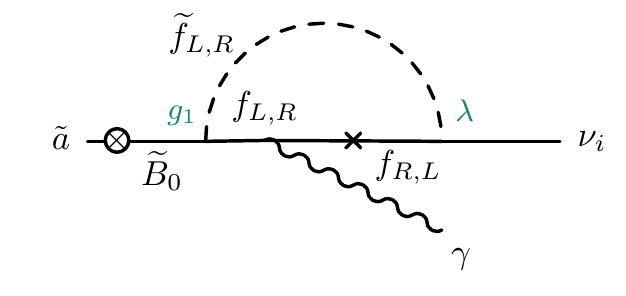}}}\\
\subfloat[]{\raisebox{5ex}{\label{fig:LoopDiagBosH}\includegraphics[width=0.45\textwidth]{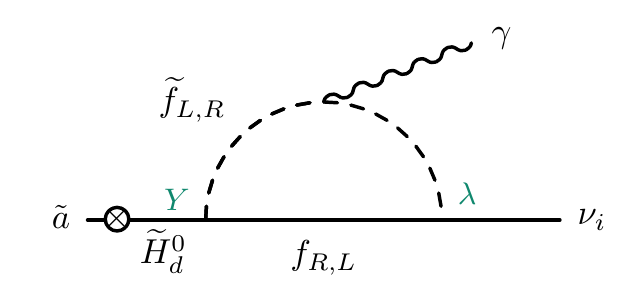}}}
\hspace{1cm}
\subfloat[][]{{\label{fig:LoopDiagFerH}\includegraphics[width=0.45\textwidth]{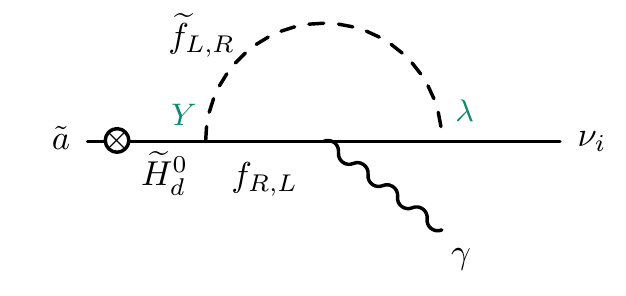}}}\\
\subfloat[]{\raisebox{5ex}{\label{fig:LoopDiagBosR}\includegraphics[width=0.45\textwidth]{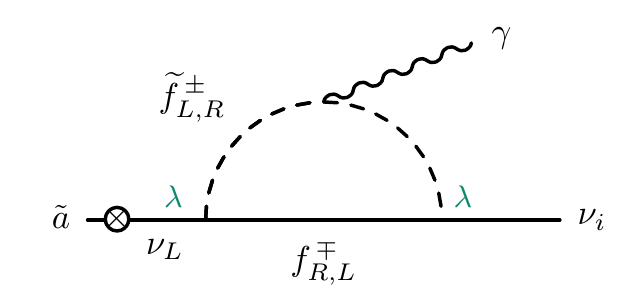}}}
\hspace{1cm}
\subfloat[][]{{\label{fig:LoopDiagFerR}\includegraphics[width=0.45\textwidth]{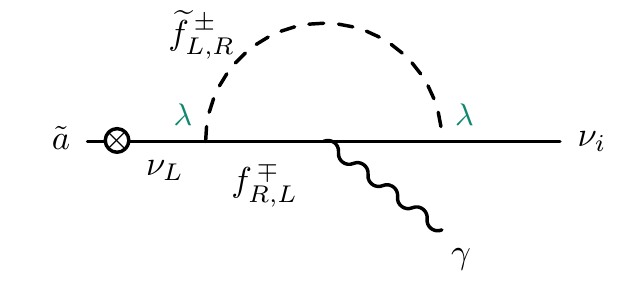}}}\\
\caption{Diagrams with an explicit dependence on the RpV trilinear couplings $\lambda^{(\prime)}$. The notation is as in Fig.~\ref{fig:Diagrams_noRpV}.} \label{fig:Diagrams_RpV}
\end{figure}

The one-loop diagrams for the decay $\widetilde a(p) \to \nu(k_1) + \gamma (k_2)$ have an amplitude with the following 
structure~\cite{Mukhopadhyaya:1999gy} dictated by gauge invariance
\begin{equation}
{\cal M} = i g_{\widetilde a \nu \gamma} \bar u(k_1) \left( P_R - \eta_\nu \eta_{\widetilde a} P_L \right) \sigma^{\mu\nu} 
k_{2\mu} \epsilon^*_\nu u(p) \, .
\end{equation}
Here $\eta_\nu$ and $\eta_{\widetilde a}$ are the signs of the mass eigenvalues of the neutrino and the axino, 
respectively, $k_2$ the photon momentum, and $\epsilon$ its polarization vector. $g_{\widetilde a \nu \gamma}$ is a 
function with the details of the loop integrals, and has the dimension of inverse mass. The decay rate then has the form
\begin{equation}
\Gamma_{\widetilde a \to \nu\gamma} = \frac{|g_{\tilde a\nu\gamma}|^2 m_{\widetilde a}^3}{16 \pi} \, .
\end{equation}
We are neglecting the final state neutrino mass. The corresponding Feynman diagrams are shown in 
Figs.~\ref{fig:Diagrams_noRpV} and \ref{fig:Diagrams_RpV}. Here we are interested in an estimate, and thus wish 
to determine which diagram(s) is (are) dominant. Thus we first consider the individual amplitudes squared, in turn, as
well as the corresponding decay width, without including interference terms. If not otherwise specified, we make use of 
the formulae provided in Ref.~\cite{Lavoura:2003xp} to compute the loop integrals. We do not include diagrams with 
sneutrino VEVs insertions (see {\it e.g.} Ref.~\cite{Dawson:1985vr}), since, as previously mentioned,  we work in 
the basis, defined at the axino mass scale, where such VEVs are zero.

The first diagrams we consider are those depicted in Fig.~\ref{fig:LoopDiagFerW}\,,~\ref{fig:LoopDiagBosW}.
The corresponding decay width has been computed in the analogous case of a sterile neutrino decaying into a 
neutrino and a photon~\cite{Pal:1981rm} (see also Ref.~\cite{Adhikari:2016bei} for a review), 
 \begin{align} \label{eq:GammaW}
\Gamma_{\nu(Wl)} & = \frac{9 \alpha G_F^2}{1024 \pi^4} x_{\tilde a,\nu}^2 \maxino^5 \approx  \frac{9 \alpha G_F^2}{1024 \pi^4} \frac{\kappa_{\rm eff}^2 v_u^2}{f_a^2} \maxino^3 \\ 
& = 3 \cdot 10^{-58} \ {\rm GeV} \  \left( \frac{\kappa_{\rm eff}}{\rm MeV} \right)^2 \left( \frac{10^{11} \ {\rm GeV}}{f_a}  \right)^2 \left( \frac{\maxino}{\rm keV}  \right)^3 \, . \nonumber
\end{align}
Here, the first subscript on $\Gamma$ indicates the state the incoming axino mixes with, while the subscripts 
in parentheses denote the particles running in the loop. We use this notation for the rest of the section. Let us anticipate that in 
most of the parameter space we consider, $\Gamma_{\nu(Wl)}$ gives the dominant contribution. It is therefore useful to 
compare the other partial widths to $\Gamma_{\nu(Wl)}$. Incidentally, a decay width of $\Gamma_{\tilde a}=3\cdot10^{-58}\,$GeV 
corresponds to a lifetime $\tau_{\tilde a}=2\cdot10^{23}\,$s and the lifetime of the universe is about $\tau_{\mathrm{univ.}}\approx4\cdot 
10^{17}\,$s.

The main contribution to the two diagrams in Fig.~\ref{fig:LoopDiagBosC},~\ref{fig:LoopDiagFerC} comes from the heaviest
lepton that can run in the loop, the $\tau$. We have 
\begin{equation} \label{eq:Gamma_Hl}
\Gamma_{(Hl)} \approx \frac{\maxino^3}{2^{11} \pi^5} \frac{\keff^2}{f_a^2} \left(\frac{B\mu}{m_{H^\pm}^2}\right)^2 Y_\tau^2   
\frac{m_\tau^2}{m_{H^\pm}^4} \ln^2 \left(\frac{m_{\tau}^2}{m_{H^\pm}^2}\right)  \,.
\end{equation}
The $m_{\tau}^2$ in the numerator comes from the mass insertion that flips the chirality in the fermion line in the loop, while 
$B\mu / m_{H^\pm}^{2}$ is an estimate 
of the mixing between the charged scalar Higgses, $B\mu$ being the bilinear parameter in the Higgs potential, see Eq.~\eqref{softLag}.  Taking $B\mu \sim m_{H^\pm}^2$ as an estimate, we find the ratio
\begin{align}
\frac{\Gamma_{(Hl)}}{\Gamma_{\nu(Wl)}}&=\frac{1}{18 \pi \alpha G_F^2}  \frac{Y_{\tau}^2}{v_u^2} \frac{m_{\tau}^2}
{m_{H^{\pm}}^4} \ln^2 \left(\frac{m_{\tau}^2}{m_{H^\pm}^2}\right) \\ \nonumber
& \approx 3 \cdot 10^{-8} \left(\frac{m_{H^{\pm}}}{\tev}\right)^{-4} \frac{ \ln^2 \left(\frac{m_{\tau}^2}{m_{H^\pm}^2}\right)}{160} \, ,
\end{align}
which is highly  suppressed, and we neglect the contributions from the corresponding diagrams in the following.

Next we consider the diagrams of Fig.~\ref{fig:LoopDiagBos2C},~\ref{fig:LoopDiagFer2C}. We assume the $\tilde \tau$ to be the lightest
charged slepton and give the leading contribution. We find
\begin{equation}
\Gamma_{(\tilde H \tilde l)} \approx \frac{\maxino^3}{2^{12} \pi^5} \frac{\keff^2}{f_a^2} x_{\tilde\tau_{LR}}^2 Y_\tau^2   
\frac{\mu_{\rm eff}^2}{m_{\tilde \tau}^4} \frac{\left(1+\ln \frac{\mu_{\rm eff}^2}{m_{\tilde\tau}^2}-\frac{\mu_{\rm eff}^2}
{m_{\tilde{\tau}}^2}\right)^2}{\left(1-\frac{\mu_{\rm eff}^2}{m_{\tilde{\tau}}^2}\right)^4} \, .
\end{equation}
Here, $x_{\tilde\tau_{LR}}$ is the mixing between the left and right staus, $\mu_{\rm eff}$ is the Higgsino 
mass. Note that the expression is finite in the limit $\mu_{\rm eff} = m_{\tilde \tau}$. Taking $x_{\tilde\tau_{L R}} 
\approx 1$, we find the ratio
\begin{align}
\frac{\Gamma_{(\tilde H \tilde l)}}{\Gamma_{\nu(Wl)}}&=\frac{1}{18 \pi \alpha G_F^2}  \frac{Y_{\tau}^2}{v_u^2} \frac{\mu_{\rm eff}^2}{m_{\tilde \tau}^4} \frac{\left(1+\ln \frac{\mu_{\rm eff}^2}{m_{\tilde\tau}^2}-\frac{\mu_{\rm eff}^2}{m_{\tilde{\tau}}^2}\right)^2}{\left(1-\frac{\mu_{\rm eff}^2}{m_{\tilde{\tau}}^2}\right)^4}  \\ \nonumber
& \approx 10^{-5} \left(\frac{\mu_{\rm eff}}{\tev}\right)^{2} \left(\frac{m_{\tilde\tau}}{\tev}\right)^{-4}  \, .
\end{align}
This is also highly suppressed and we neglect it in the following.

Next, we examine the diagrams of Fig.~\ref{fig:Diagrams_RpV}, which involve at least one trilinear RpV coupling, $\lambda$ or 
$\lambda^{\prime}$. To avoid cluttering the equations, we use $\lambda$ to denote either. 
For the diagrams of Fig.~\ref{fig:LoopDiagBosB}\,,~\ref{fig:LoopDiagFerB} we find
\begin{equation}   \label{eq:Gamma1}
\Gamma_{\widetilde{B}\left(f\tilde{f}\right)} \approx \frac{\maxino^3 Q^2_{f}}{2^{11} \pi^5} g_1^2 \lambda^2 \mixB^2  
\frac{m_f^2}{m_{\widetilde f}^4} \ln^2 \left(\frac{m_f^2}{m_{\widetilde f}^2}\right) \, .
\end{equation}
Here, $f$ ($\widetilde f$) denotes either a down-type quark (squark), in the case of a $\lam'$ coupling, or a lepton (slepton),
in the case of a $\lam$ coupling. $Q_f$ is the electric charge of $f$ in units of $e$, and $g_1 = e / \cos \theta_W$ is the 
hypercharge gauge coupling. The dominant contributions come from loops of bottom-sbottom, or tau-stau; the two are of the 
same order of magnitude, under the assumption that sleptons and squarks have comparable masses, as the smaller mass of 
$\tau$ is compensated by its larger electric charge. The axino-bino mixing is estimated in Eq.~\eqref{abimix}. Comparing this 
contribution to the one of Eq.~\eqref{eq:GammaW}, we find
\begin{align}
\frac{\Gamma_{\widetilde{B}\left(f\tilde{f}\right)}}{\Gamma_{\nu\left(Wl\right)}}&=\frac{g_1^4 Q^2}{18\pi \alpha G_F^2}
\frac{\lambda^2 v^2}{M_{\rm SUSY}^2 \keff^2}   \frac{m_f^2}{m_{\widetilde f}^4}\ln^2\left( \frac{m_f^2}{m_{\widetilde f}^2}\right)   
\label{RatioGB} \\ \nonumber
&\approx 0.2  \left( \frac{\lambda}{10^{-2}} \right)^2 \!\left( \frac{Q_f}{1/3} \right)^2 \left( \frac{\mev}{\keff}
\right)^{2} \left( 
\frac{\tev}{ M_{\rm SUSY}}\right)^{2}  
\left( \frac{m_f}{m_b} \right)^2 \left( \frac{\tev} {m_{\widetilde f}}
\right)^4 \frac{\ln^2\left( 
\frac{m_f^2}{m_{\widetilde f}^2}\right)}{120} \,,
\end{align}
where $m_b$ is the mass of the bottom quark. For sfermions lighter than a TeV, $\Gamma_{\widetilde{B}\left(f\tilde{f}\right)}$
would give a contribution larger than $\Gamma_{\nu\left(Wl\right)}$. Here we restrict ourselves to $m_{\widetilde f} > 1$ TeV, 
motivated by recent analyses~\cite{CMS:2017mkt, Dercks:2017lfq}, which disfavor lighter squarks and sleptons. We 
comment further on this contribution below.

For the diagrams of Fig.~\ref{fig:LoopDiagBosH}\,,~\ref{fig:LoopDiagFerH} we find (note there is no mass insertion in the 
fermion line in the loop)
\begin{equation}   \label{eq:gamma2}
\Gamma_{\widetilde{H}\left(f\tilde{f}\right)} \approx \frac{\maxino^3 Q^2_f}{2^{11} \pi^5} Y^2 \lambda^2 \mixH^2 
\frac{\maxino^2}{m_{\widetilde f}^4} \ln^2 \left( \frac{m_f^2}{m_{\widetilde f}^2}\right)  \, .
\end{equation}
Here, $f$ ($\widetilde f$) can be either a down-type quark (squark), in the case of a $\lam'$ coupling, or a lepton (slepton), 
in the case of a $\lam$ coupling. $Y$ is either the $Y_d$ or the $Y_e$ Yukawa, and the axino-higgsino mixing is 
given in Eq.~\eqref{ahimix}. As in the previous case, the main contributions to these diagrams are from bottom-sbottom and 
tau-stau loops, respectively.
 $\Gamma_{\widetilde{H}\left(f\tilde{f}\right)}$ is suppressed compared to $\Gamma_{
\widetilde{B}\left(f\tilde{f}\right)}$ by a factor $\maxino^2 / m_f^2$. We have
\begin{align}
\frac{\Gamma_{\widetilde{H}\left(f\tilde{f}\right)}}{\Gamma_{\nu\left(Wl\right)}} &= \frac{Q^2_f}{18 \pi\alpha G_F^2} 
\frac{\lambda^2 Y^2}{\keff^2} \frac{\maxino^2}{m_{\widetilde f}^4} \ln^2\left( \frac{m_f^2}{m_{\widetilde f}^2}\right) \\ \nonumber
&\approx 10^{-12} \left( \frac{\lambda}{10^{-2}}\right)^2 \left( \frac{Q_f}{1/3}\right)^2 \left( \frac{Y}{Y_b} \right)^2 
\left(\frac{\mev}{\keff}
\right)^{2} \left(\frac{\maxino}{10 \ \kev}\right)^2  \left( 
\frac{\tev}{m_{\widetilde f}}
 \right)^{4} 
\frac{ \ln^2 \left( \frac{m_f^2}{m_{\widetilde f}^2}\right) }{120} \, .
\end{align}

\begin{figure}[t]
\centering
\includegraphics[width=0.7\textwidth]{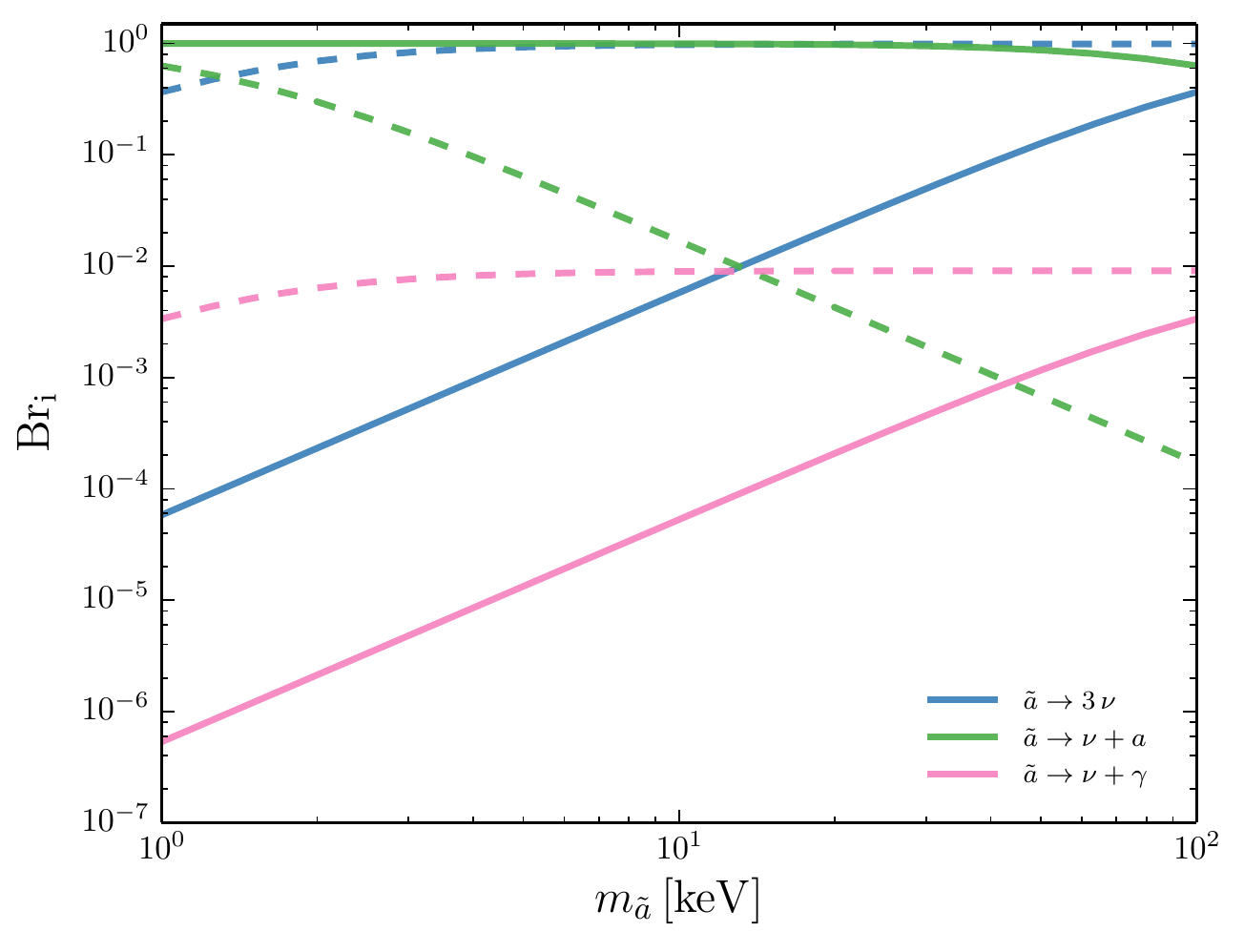} 
\caption{We show the branching ratios of the axino decay channels into axion and neutrino, three neutrinos, photon
and neutrino, as a function of the axino mass. The total width is $\Gamma_{\tilde a \to a \nu} + \Gamma_{\tilde a \to 3 \nu} +
\Gamma_{\tilde a \to \gamma \nu}$, where in $\Gamma_{\tilde a \to \gamma \nu}$ we take into account only the two dominant contributions, $\Gamma_{\widetilde{B}\left(f\tilde{f}\right)} + \Gamma_{\nu (Wl)}$, as discussed in the main text.
We consider a loop of bottom-sbottom for the evaluation of $\Gamma_{\widetilde{B}\left(f\tilde{f}\right)}$, with $m_{\tilde b} = 1$ TeV (the tau-stau contribution would be analogous) and the RpV trilinear coupling $\lambda = 10^{-2}$. 
 Continuous (dashed) lines are for $f_a = 10^{10}$ GeV ($f_a = 10^{12}$ GeV).} \label{fig:BRatios}
\end{figure}

The last two diagrams to evaluate are those of Fig.~\ref{fig:LoopDiagBosR},~\ref{fig:LoopDiagFerR}. The two RpV 
couplings appearing in each of these diagrams have no restrictions on their flavor index structure, contrary to those 
of~\crefrange{fig:LoopDiagBosB}{fig:LoopDiagFerH}, which are diagonal in the singlet-doublet mixings. For the sake of our 
discussion we also do not distinguish between the two potentially different $\lambda$ here, as we consider only one RpV 
coupling to be different from zero at a time. We find
\begin{equation}
\Gamma_{\nu\left(f\tilde{f}\right)} \approx \frac{\maxino^3 Q^2_f}{2^{11} \pi^5} \lambda^4 \mixN^2  \frac{\maxino^2}{m_{\widetilde f}^4} \ln^2 \left(\frac{m_f^2}{m_{\widetilde f}^2}\right)  \, ,
\end{equation}
the main contributions being again from bottom-sbottom and tau-stau loops, and
\begin{align}
\frac{\Gamma_{\nu\left(f\tilde{f}\right)}}{\Gamma_{\nu\left(Wl\right)}}  &= \frac{Q^2_f}{18 \pi \alpha G_F^2} \frac{\lambda^4}{m_{\widetilde f}^4} \left(\log \frac{m_f^2}{m_{\widetilde f}^2}\right)^2 \\ \nonumber
&\approx 10^{-9} \left( \frac{\lambda}{10^{-2}}\right)^4 \left( \frac{Q_f}{1/3}\right)^2 \left( \frac{\tev} {m_{\widetilde f}}
\right)^{4} \frac{ \ln^2\left( \frac{m_f^2}{m_{\widetilde f}^2}\right) }{120} \, .
\end{align}
From our estimates we see that $\Gamma_{\widetilde{B}\left(f\tilde{f}\right)}$ and $\Gamma_{\nu (Wl)}$ give the dominant 
contributions,
the other diagrams being always negligible. 
Thus the axion decay width into photon and neutrino is given to a good approximation by 
$\Gamma_{\widetilde{B}\left(f\tilde{f}\right)} + \Gamma_{\nu (Wl)}$. 
In Fig.~\ref{fig:BRatios} we plot the branching ratios corresponding to the three decay channels of our light axino.

\begin{figure}[t]
\centering
\subfloat{\includegraphics[width=0.45\textwidth]{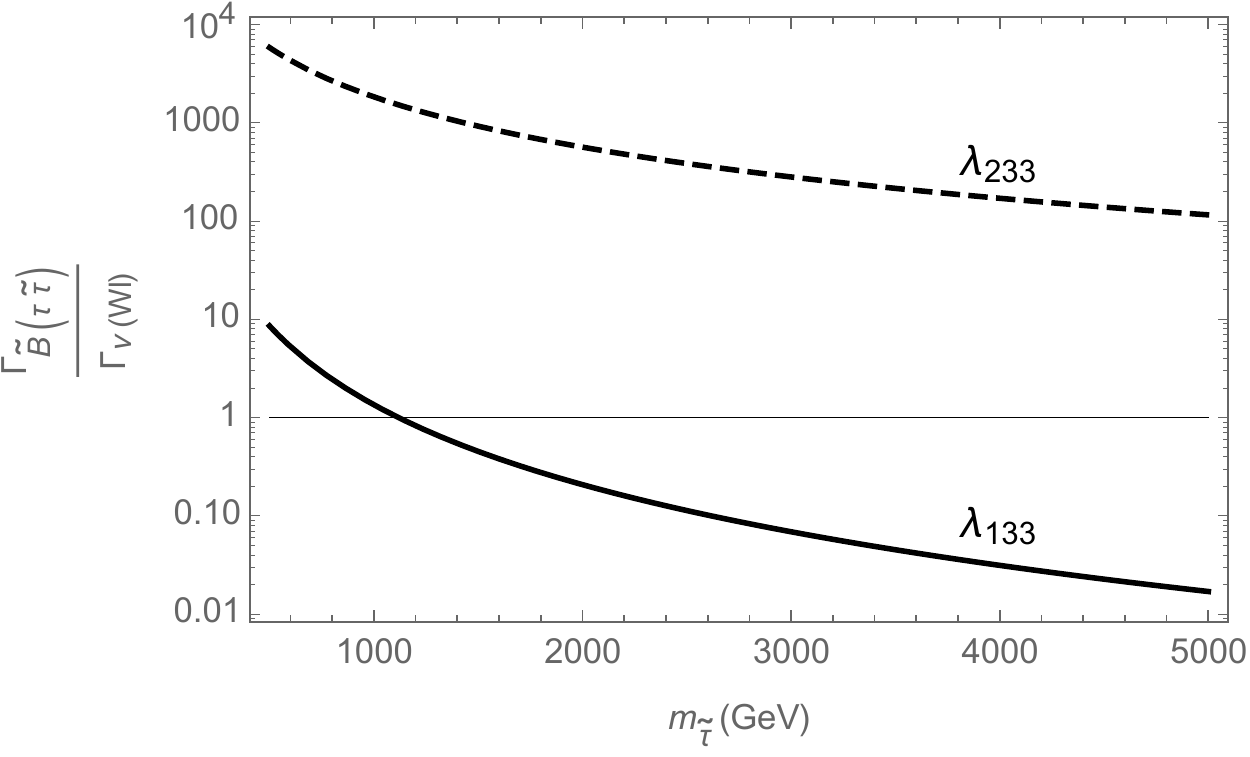}} \qquad
\subfloat{\includegraphics[width=0.45\textwidth]{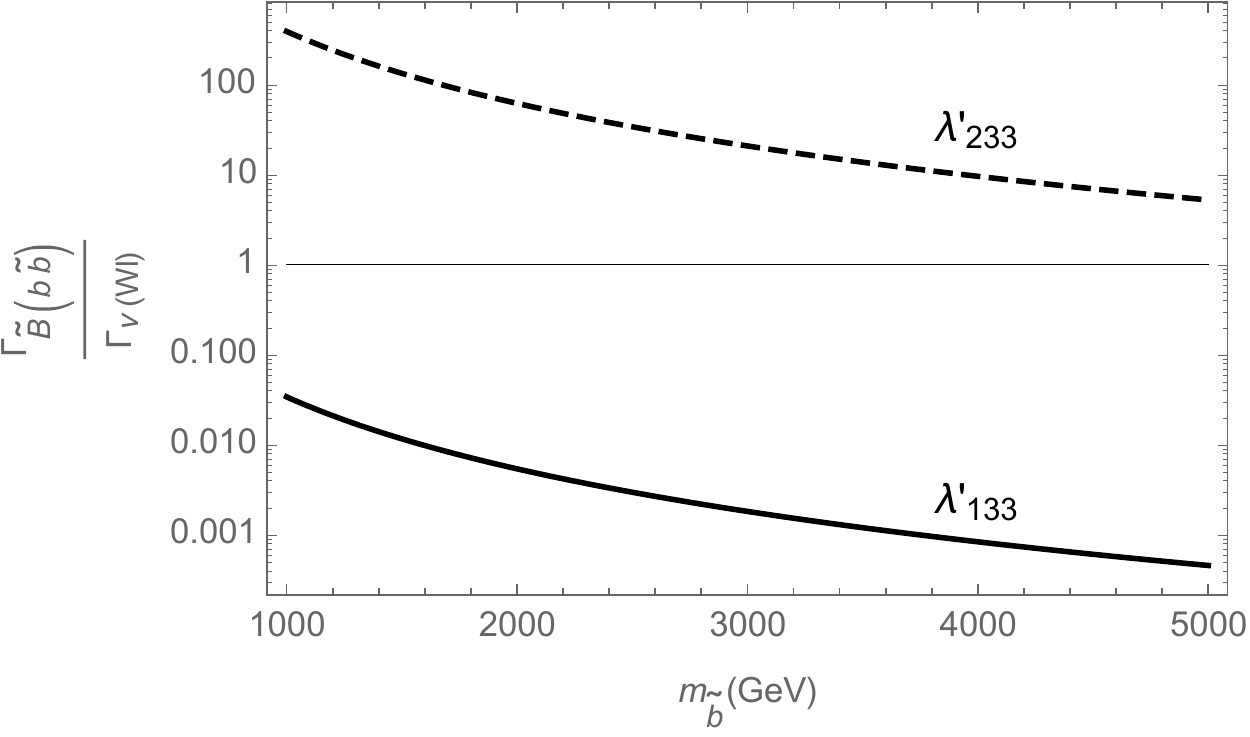}}
\caption{We show the ratio $\frac{\Gamma_{\widetilde{B}\left(f\tilde{f}\right)}}{\Gamma_{\nu\left(Wl\right)}}$ from Eq.~\eqref{RatioGB} as a function of the sfermion mass. We have set $M_{\rm SUSY} = 1$ TeV and $\keff = 1$ MeV. 
The left plot corresponds to the tau-stau loop, with $m_f = m_\tau$. The solid line is obtained by taking
$\lambda = \lambda_{133} = 0.006 \sqrt{\frac{m_{\tilde \tau}}{100 \ {\rm GeV}}}$, while for the dashed line
$\lambda = \lambda_{233} = 0.07 \frac{m_{\tilde \tau}}{100 \ {\rm GeV}}$.
The right plot corresponds to the bottom-sbottom loop, with $m_f = m_b$. The solid line is obtained by taking
$\lambda^{'} = \lambda^{'}_{133} = 0.0014 \sqrt{\frac{m_{\tilde b}}{100 \ {\rm GeV}}}$, while for the dashed line
$\lambda = \lambda^{'}_{233} = 0.15 \sqrt{\frac{m_{\tilde b}}{100 \ {\rm GeV}}}$. The values chosen for the trilinear
couplings saturate the up-to-date bounds of Table~\ref{tab:bounds1}.
} \label{fig:Ratiolam}
\end{figure}

It is worth examining $\Gamma_{\widetilde{B}\left(f\tilde{f}\right)}$ in more detail. First, we reintroduce the generation
indices in the superpotential terms 
\be
\frac{1}{2} \lambda_{ijk} \hat L_i \hat L_j \hat {\bar E}_k + \lambda^{'}_{ijk} \hat L_i \hat Q_j \hat {\bar D}_k \, .
\ee
 As the main contributions to $\Gamma_{\widetilde{B}\left(f\tilde{f}\right)}$ are from tau-stau and bottom-sbottom loops, the trilinear couplings of interest are those shown in Table~\ref{tab:bounds1}. 
The first index of $\lambda_{ijk}$ and $\lambda^{'}_{ijk}$ refers, in our case, to the neutrino in the lepton doublet. 
We see in Fig.~\ref{fig:Ratiolam} that when we consider the decay $\widetilde a \to \nu_2 \gamma$, with 
$\lambda_{233}$ and $\lambda^{'}_{233}$ saturating the upper bounds in Table~\ref{tab:bounds1}, the diagrams giving
$\Gamma_{\widetilde{B}\left(f\tilde{f}\right)}$ provide the dominant contribution. 

\begin{figure}[t]
\centering
\subfloat{\includegraphics[width=0.45\textwidth]{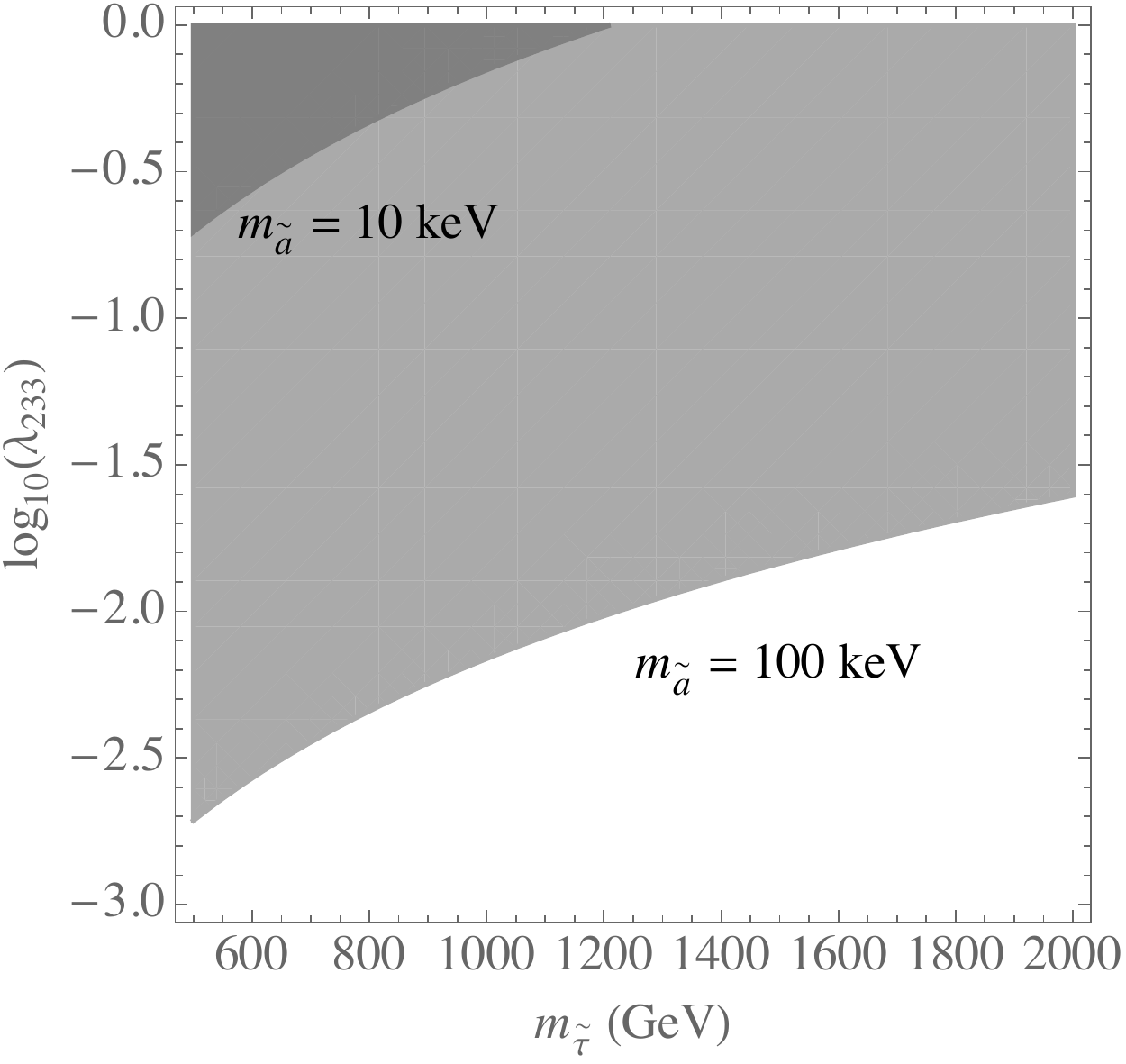}} \qquad
\subfloat{\includegraphics[width=0.45\textwidth]{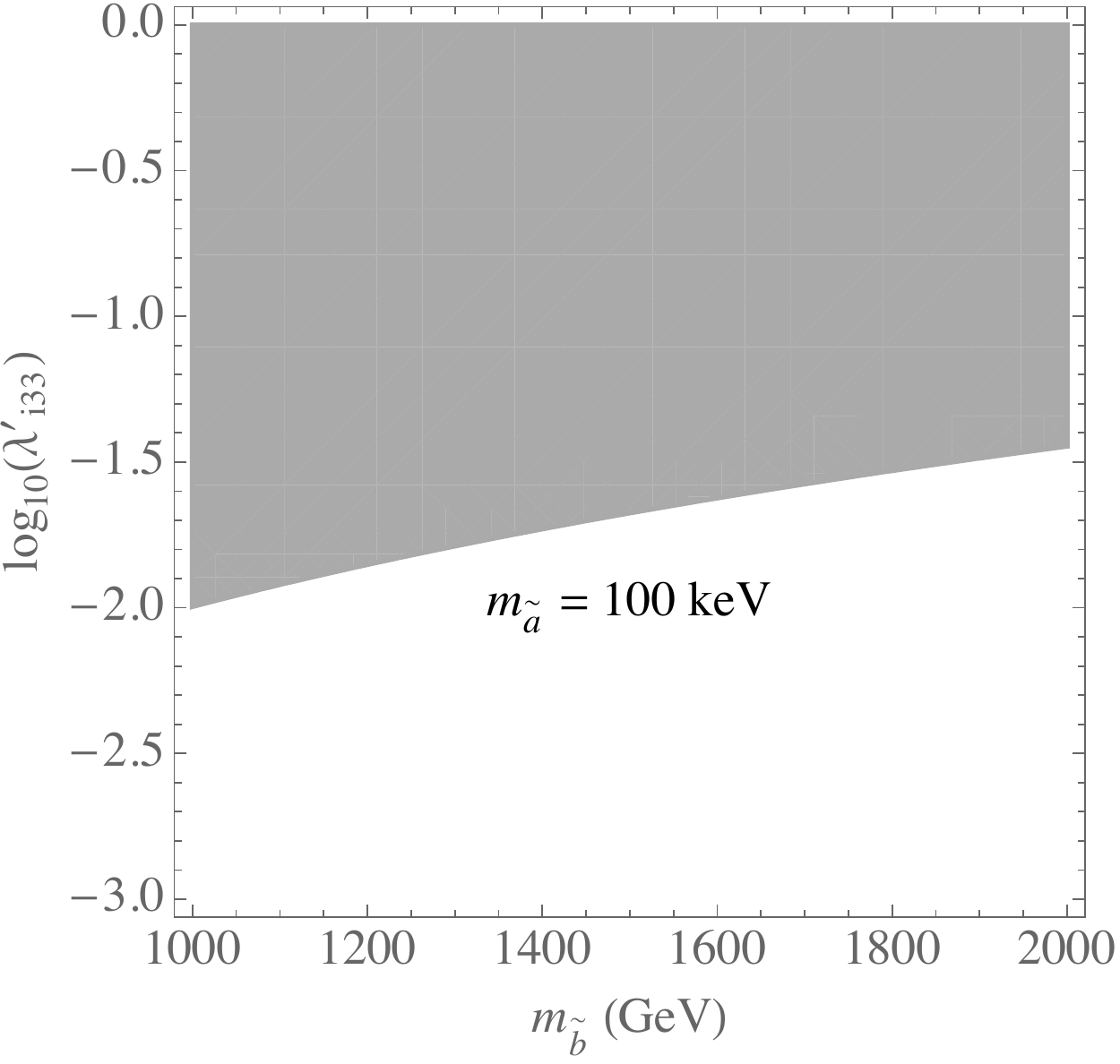}}
\caption{We show regions excluded by X and gamma-ray constraints~\cite{Arias:2012az, Essig:2013goa}. 
Here we use Eq.~\eqref{Xbound} and assume
the axino to contribute to the totality of dark matter. The dark (light) gray region is excluded when $\maxino = 10$ keV (100 keV). In $\lambda^{'}_{i33}$, the index $i$ can be either 2 or 3, in which case the dominant contribution
to the decay $\widetilde a \to \nu_i \gamma$ is from $\Gamma_{\widetilde{B}\left(f\tilde{f}\right)}$.
}
 \label{fig:lambounds}
\end{figure}

We can use this fact to set new bounds on these trilinear couplings by considering constraints from X and gamma 
rays~\cite{Arias:2012az, Essig:2013goa} on the decays $\widetilde a \to \nu_{2,3} \gamma$. We proceed as follows. First, 
we assume that the entire dark matter is constituted by the axino, $\Omega_{\widetilde a} h^2 = \Omega_{\rm DM}$. This 
fixes $f_a$ for any given value of $\maxino$. Then the excluded region corresponds to
\be \label{Xbound}
\Gamma_{\widetilde{B}\left(f\tilde{f}\right)} > \Gamma_{\rm bound} = \frac{1}{\tau_{\rm bound}} \, ,
\ee
where $\tau_{\rm bound}$ is given, for instance, in Fig.~9 of Ref.~\cite{Essig:2013goa} as a function of the mass of the 
decaying particle, the axino in our case.  In Fig.~\ref{fig:lambounds} we see that, for $\maxino = 100$ keV, we can set 
bounds on the trilinears $\lambda_{233}, \lambda^{'}_{233}, \lambda^{'}_{333}$ of order $10^{-2}$. 
These are to be compared with the bounds in Table~\ref{tab:bounds1}. However, it is important to keep in mind that these new bounds,
contrary to those in Table~\ref{tab:bounds1}, rely on the presence of the axino and on the assumption that it is the whole 
dark matter.
\begin{table*}
\centering
\renewcommand\arraystretch{1.4}
\begin{tabular*}{\textwidth}{l @{\extracolsep{\fill}} cc}
\toprule
ijk &\(\lambda_{ijk}(M_W)\)&\(\lambda^\prime_{ijk}(M_W)\) \\ &&\\[-5mm]
\hline &&\\[-5mm]
133 & \(0.0060\times\sqrt{\frac{m_{\tilde{\tau}}}{\SI{100}{GeV}}}\) & \(0.0014\times\sqrt{ \frac{m_{\tilde{b}}}{\SI{100}{GeV}}}\) \\
233 & \(0.070\times\frac{m_{\tilde{\tau}_R}}{\SI{100}{GeV}}\) & \(0.15\times\sqrt{ \frac{m_{\tilde{b}}}{\SI{100}{GeV}}}\) \\
333 & - & \(0.45\ (1.04)\) \\
\bottomrule
\end{tabular*}
\caption{Upper bounds on the magnitude of R-parity violating couplings at the $2\sigma$ confidence level, taken from 
Ref.~\cite{Dercks:2017lfq}. The constraints arise from indirect decays. The concrete processes are described in detail in 
Ref.~\cite{Allanach:1999ic}. }
\label{tab:bounds1}
\end{table*}

In Fig.~\ref{fig:Money} we show the X and gamma-ray constraints on the plane $f_a$ vs $m_a$. The excluded regions
in purple correspond to
\be \label{XboundDM}
\frac{\Omega_{\widetilde a}}{\Omega_{\rm DM}}\Gamma_{\widetilde a \to \nu \gamma} > \Gamma_{\rm bound} = \frac{1}{\tau_{\rm bound}} \, .
\ee
Here, $\Gamma_{\widetilde a \to \nu \gamma} = {\rm max}[\Gamma_{\widetilde{B}\left(f\tilde{f}\right)}, \Gamma_{\nu \left(Wl\right)}]$, and we multiply by $\Omega_{\widetilde a} / \Omega_{\rm DM}$ to account for the region of parameter space
where the axino is under abundant. We discuss the axino relic abundance in Sec.~\ref{sec:cosmology}.

\subsection{Comment on the 3.5 keV Line.}
This model can fit the 3.5 keV line observed in galaxy clusters~\cite{Bulbul:2014sua, Boyarsky:2014jta, Liew:2014gia, Kong:2014gea}.
A possible benchmark point, for instance, is the following:
\be \label{bench35}
\maxino = 7 \ {\rm keV} \, , \ f_a = 3.5 \cdot 10^{10} \ {\rm GeV} \, , \ m_{\widetilde b} = 2 \ {\rm TeV} \, , \ \lambda^{\prime}_{233} = 0.22 \, .
\ee
With these values, the axino constitutes the entire dark matter 
and has a partial decay width $\Gamma_{\widetilde a \to \nu \gamma}  \approx \Gamma_{\widetilde{B}\left(f\tilde{f}\right)} \approx
6 \cdot 10^{-53}$ GeV, needed to explain the putative line. The benchmark complies with the bounds we mentioned above.
We emphasize, however, that evidence that the 3.5 keV line is due to decaying DM 
 is not conclusive~\cite{Urban:2014yda, Jeltema:2014qfa, Bulbul:2014ala, Boyarsky:2014paa, Jeltema:2014mla,Carlson:2014lla, Ruchayskiy:2015onc, Jeltema:2015mee, Bulbul:2016yop}.

\section{Cosmological and Astrophysical Constraints} \label{sec:cosmology}
%
\subsection{The Gravitino}
We assume a supergravity completion of our model, with SUSY broken at a high scale ($\sqrt{F}\sim 3\cdot10^{10}$ GeV) and the breaking effects
mediated via Planck suppressed operators, so that the soft scale is $M_{\rm SUSY} \sim \frac{F}{M_{\rm Pl}} = {\cal O}$(TeV).
The gravitino is heavy, with $m_{3/2} \sim M_{\rm SUSY}$, decays before BBN, and poses no cosmological problems.

\subsection{The Saxion}
In the context of supersymmetric axion models one usually has to worry about the saxion: if it is too light it can pose serious issues (see \textit{e.g.} Ref.~\cite{Heckman:2008jy}).
 We briefly review why. The saxion is a pseudomodulus, meaning that its potential is quite flat. During inflation we expect it to sit at a distance of order 
 $f_a$ from the minimum of its zero-temperature potential. After inflation, when the Hubble parameter $H$ becomes comparable to the saxion mass, $m_s$,
 the saxion starts to oscillate about its minimum. From the condition $H(T_s^{\rm osc}) \sim m_s$, with $H(T) \sim \frac{T^2}
 {M_{\rm Pl}}$, we find the temperature at which the oscillations start
\be
T_s^{\rm osc} \sim \sqrt{m_s M_{\rm Pl}} \, .
\ee
At that time the energy density in the oscillating field is $\rho_s \sim m_s^2 f_a^2$. The energy density of radiation is 
$\rho_r \sim (T_s^{\rm osc})^4$, so that $\left( \frac{\rho_s}{\rho_r} \right)_{\rm osc} \sim \frac{f_a^2}{M_{\rm Pl}^2}$. 
The energy density of the oscillating saxion scales as $T^3$, behaving like matter, which implies that in the absence 
of decays it would come to dominate the energy density of the universe at a temperature
\be
T_s^{\rm dom} \sim \frac{f_a^2}{M_{\rm Pl}^2} T_s^{\rm osc} \sim \frac{f_a^2}{M_{\rm Pl}^2}\sqrt{m_s M_{\rm Pl}} \, .
\ee
It is important to determine whether it decays before or after domination. To proceed with this simple estimate we 
parametrize the decay rate as $\Gamma_s \sim \frac{1}{16 \pi} \frac{m_s^3}{f_a^2}$, neglecting the masses of the decay 
products. The factor of $f_a^2$ at the denominator is due to the fact that the saxion couplings to matter are suppressed by 
$f_a$. Then the decay temperature is obtained from $H(T_s^{\rm dec}) \sim \Gamma_s$:
\be
T_s^{\rm dec} \sim \frac{m_s^{3/2} M_{\rm Pl}^{1/2}}{10 f_a} \, .
\ee
From these estimates we find 
\be
\frac{T_s^{\rm dec}}{T_s^{\rm dom}} \sim 10^4 \left( \frac{m_s}{1 \ {\rm TeV}} \right) \left( \frac{10^{12} \ {\rm GeV}}{f_a}  \right)^3 \, .
\ee
In our model the saxion mass is of order TeV, so it decays way before it comes to dominate the energy density of the 
universe.\footnote{
A saxion that comes to dominate the energy density can have other interesting implications, 
see {\it e.g.} Refs.~\cite{Co:2016vsi, Co:2017orl}.} 
Note the temperature when it decays is of order GeV, safely above BBN.

\subsection{Axion Dark Matter}
In this model the axion is a dark matter candidate.
We do not review here the calculations of its relic abundance, but refer the reader to Ref.~\cite{Kawasaki:2013ae, Bae:2011jb}. 
There are two important mechanisms to produce axion cold DM. One is the misalignment mechanism, which results in the abundance
\be \label{eq:mis}
\Omega_{a, \rm mis} h^2 = 0.18 \ \theta_1^2 \left(\frac{f_a}{10^{12} \ {\rm GeV}} \right)^{1.19} \left(\frac{\Lambda_{\rm QCD}}{400 \ {\rm MeV}} \right) \, ,
\ee
with $\theta_1$ the initial misalignment angle.
The other contribution comes from axions emitted by global axionic strings, which also contribute as non-relativistic axions:
\be \label{eq:string}
\Omega_{a, {\rm str}} h^2 \approx 0.2 \ \xi \ \bar{r} \left(\frac{f_a}{10^{12} \ {\rm GeV}} \right)^{1.19} \left(\frac{\Lambda_{\rm QCD}}{400 \ {\rm MeV}} \right) \, .
\ee
Here $\xi$ is the length parameter which represents the average number of strings in a horizon volume. It is determined by 
numerical simulations~\cite{Hiramatsu:2010yu}, $\xi \simeq 1.0$. The parameter $\bar{r}$ is related     to the average energy
of the axions emitted in string decays. Its value has been the subject of a long debate. Two scenarios have been put forth, one
predicts $\bar{r} \approx 1$, the other $\bar{r} \approx 70$. See Ref.~\cite{Kuster:2008zz} and references therein for
details, and Ref.~\cite{Gorghetto:2018myk} for recent considerations on the issue.   

While the misalignment mechanism produces axions when the temperature is $T \sim \Lambda_{\rm QCD}$, the strings form at a much 
higher temperature $T \sim f_a$.  If the PQ phase transition happened before the end of inflation, or equivalently if the reheating 
temperature ($T_{\rm RH}$) were lower than $f_a$, then the strings would be inflated away and we would no longer have a contribution 
corresponding to Eq.~\eqref{eq:string}. In this scenario the axion abundance is from Eq.~\eqref{eq:mis}, and $\theta_1$ can take 
any value between $-\pi$ and $\pi$. In the scenario with the PQ phase transition post inflation ($T_{\rm RH} > f_a$), one has to average 
$\theta_1^2$ over many QCD horizons, $\langle \theta_1^2 \rangle = \pi^2/3$, and the contribution from string decays can be comparable 
to that from misalignment or larger, depending on the value of $\bar{r}$. 

\subsection{A Heavy Axino}
For $\lambda_\chi \sim 1$, we have $m_{\widetilde a} \sim v_\chi \sim M_{\rm SUSY}$,
an axino mass of order TeV. Such an axino has many open decay channels into MSSM particles, 
with a total width that can be estimated as
$\Gamma _{\widetilde a} \sim \frac{1}{8\pi} \frac{m_{\widetilde a}^3}{f_a^2}$. It decays safely before BBN, and does not pose cosmological issues.

\subsection{A Light Axino}
\begin{figure}[t!]
\includegraphics[width=0.45\textwidth]{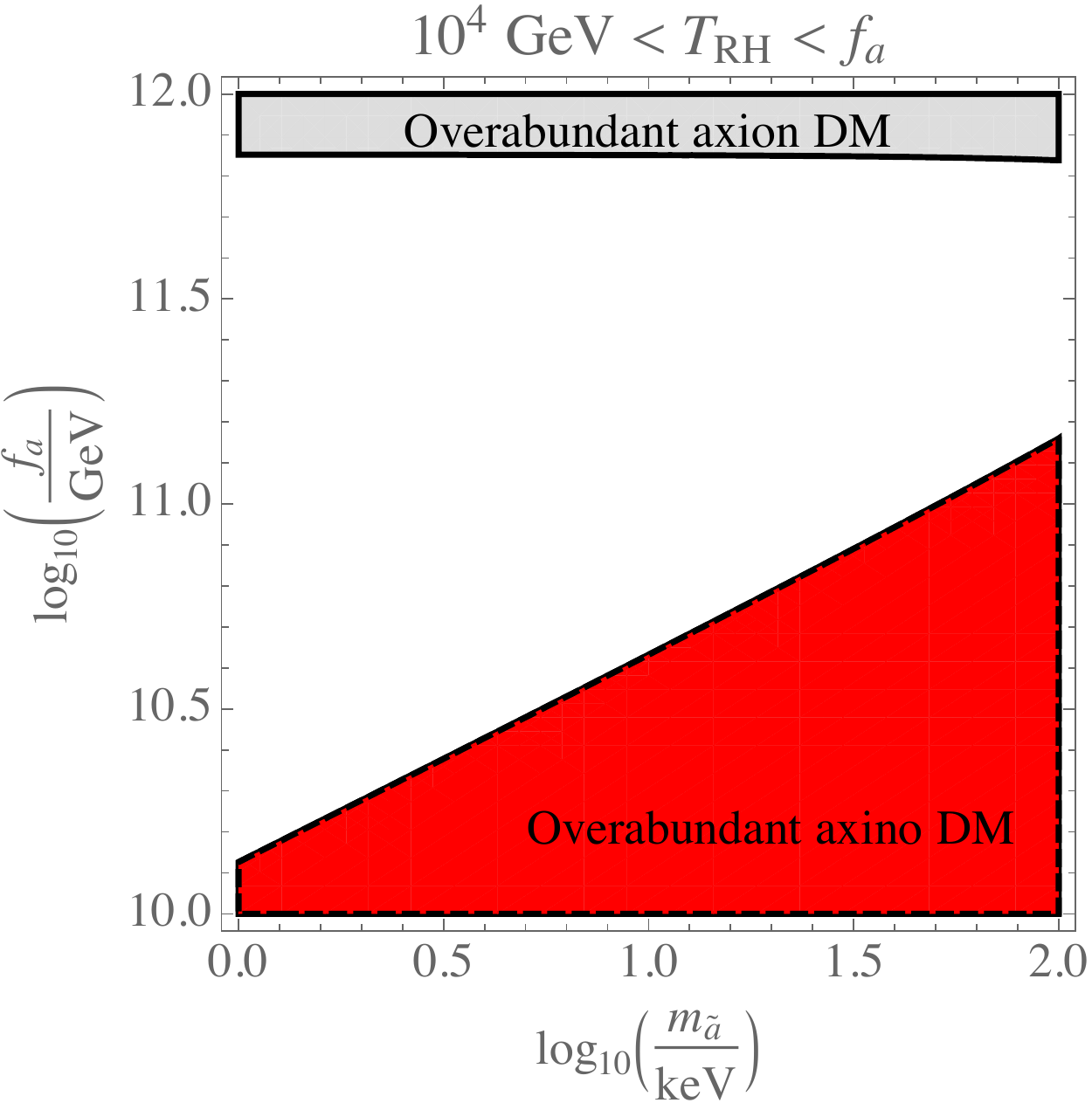}
\hspace{0.05\textwidth}
\includegraphics[width=0.45\textwidth]{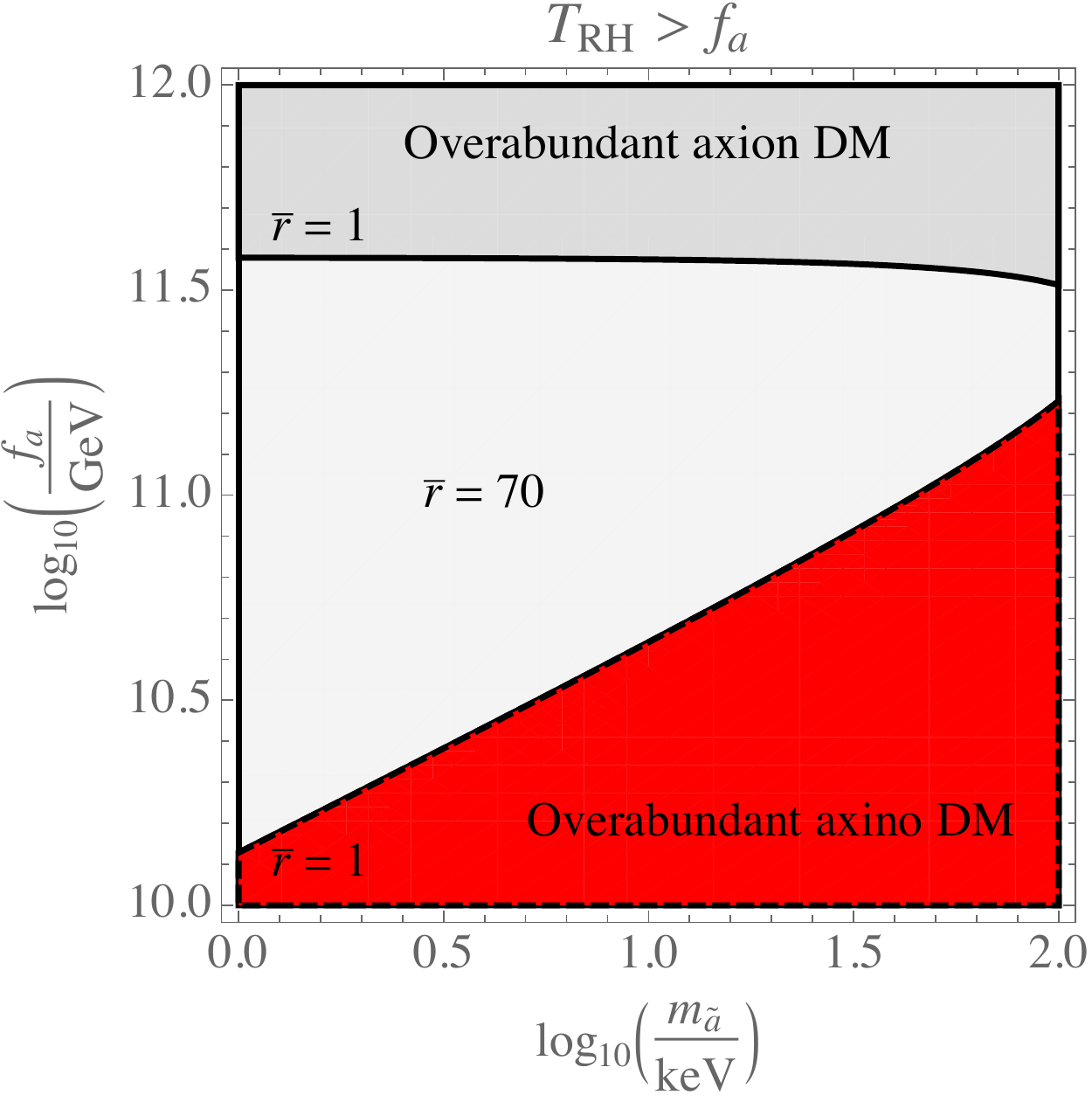}
\caption{
We show, in the plane $f_a$ vs the axino mass $m_{\widetilde a}$, the regions where the combination of axion and the axino dark 
matter exceeds the observed abundance, $(\Omega_a + \Omega_{\widetilde a}) h^2 > \Omega_{\rm DM}h^2 = 
0.12$~\cite{Ade:2015xua}. In the grey region the main contribution is from the axion, in the red from the axino [Eq.~\eqref{eq:axinoabundance}]. Along the black 
solid line the axion alone results in the correct abundance, while along the black dashed line the axino alone accounts for $\Omega_
{\rm DM}$. In the left panel the reheat temperature, $T_{\rm RH}$, is below the PQ phase transition scale, $f_a$, and the only 
contribution to axion dark matter is from the misalignment mechanism, Eq.~\eqref{eq:mis}. Here we take the initial misalignment 
angle $\theta_1 = 1$.  In the white region, the combination of axion plus axino results in a total abundance below $\Omega_{\rm DM}$. 
In the right panel $T_{\rm RH}$ is above $f_a$, and we also have a contribution to axion dark matter from the decay of axionic strings, 
Eq.~\eqref{eq:string}. We show the excluded region for two different values of the parameter $\bar r$. Note that for $\bar r = 70$, axion 
plus axino are overabundant on the entire plane. Furthermore, the lifetime of the axino is longer than the age of the universe in the 
entire region plotted.}
\label{fig:DM}
\end{figure}

\begin{figure}[t]
\includegraphics[width=0.45\textwidth]{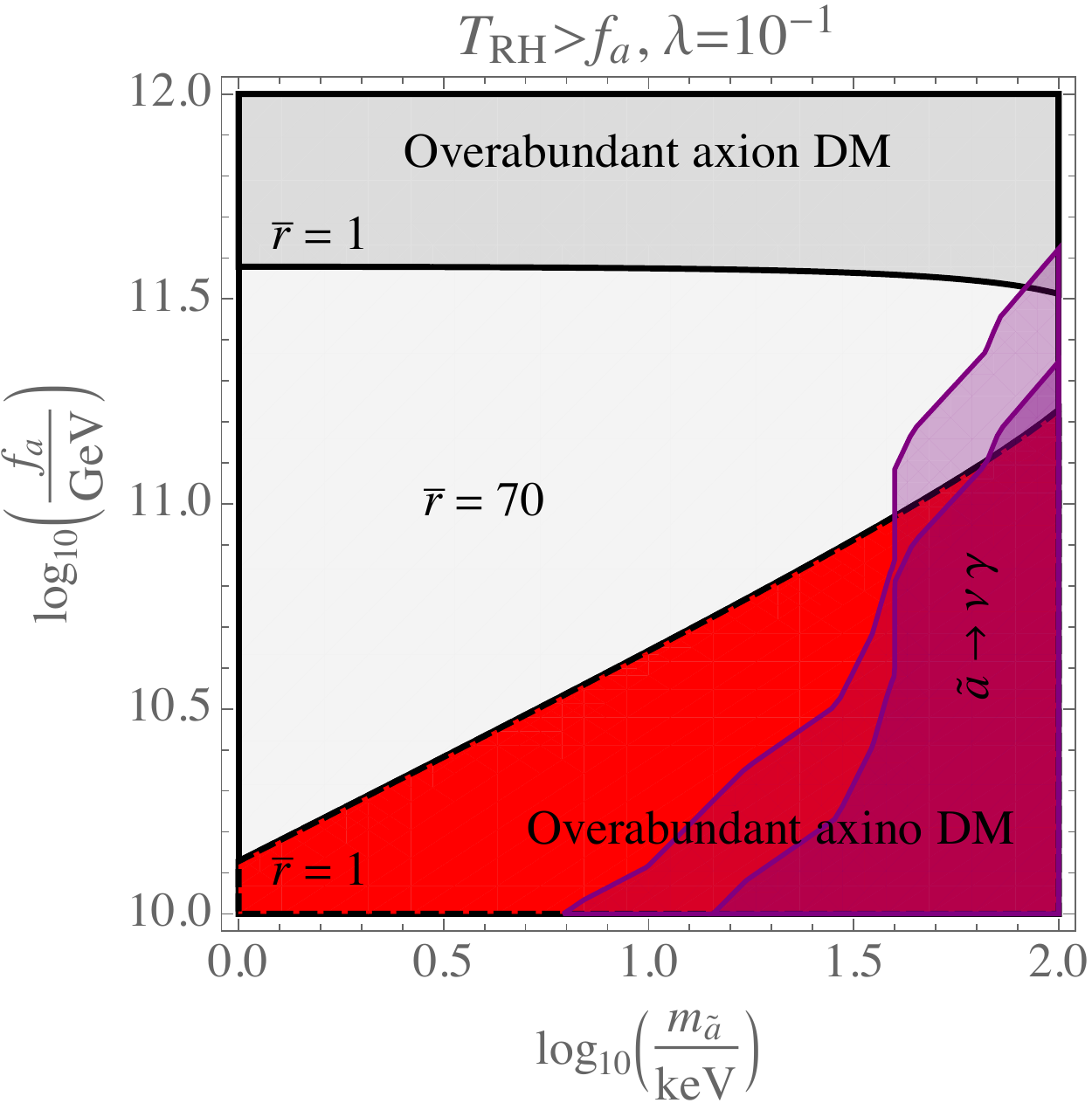}
\hspace{0.05\textwidth}
\includegraphics[width=0.45\textwidth]{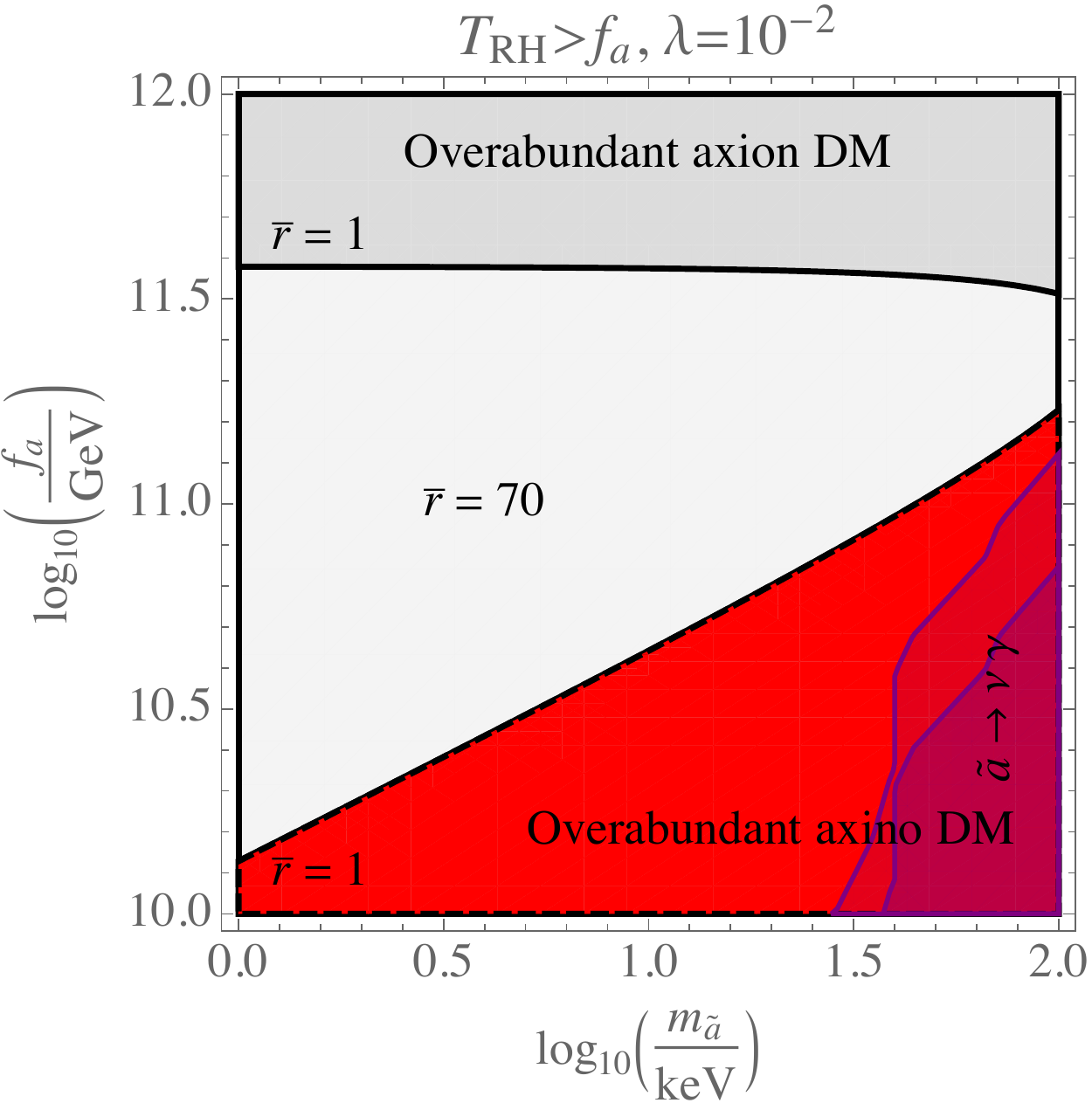}
\caption{ We add to the right plot of Fig.~\ref{fig:DM} the constraints from X and gamma rays [see Eq.~\eqref{XboundDM}], 
shown in light purple. 
When these regions overlap with the red region, the parameter space is excluded by the overabundance of the axino. 
On the left we set the trilinear coupling to $\lambda = 10^{-1}$, on the right $\lambda = 10^{-2}$. The lighter (darker) purple
region corresponds to a mass of the sfermion, running in the loop, of 1 (2) TeV. 
}
   \label{fig:Money}
\end{figure}

The axino, if light, can also be a dark matter 
candidate~\cite{Rajagopal:1990yx,Covi:2001nw, Covi:2009pq, Strumia:2010aa, Choi:2011yf,Choi:2013lwa, Co:2016fln}. It was pointed 
out in Ref.~\cite{Rajagopal:1990yx} that for $T_{\rm RH} > f_a$ a stable axino should not exceed the mass of 1 or 2 keV, otherwise it 
would result in overabundant DM. More recently Bae et al.~\cite{Bae:2011jb, Bae:2011iw} showed that the axino production rate is 
suppressed if $M_\Phi \ll f_a$, where $M_\Phi$ is the mass of the heaviest PQ--charged and gauge--charged matter supermultiplet in 
the model. This suppression is significant in DFSZ models, where $M_\Phi$ corresponds to the Higgsino mass. The consequence of 
such a suppression is that the upper bound of 2 keV on the axino mass quoted in Ref.~\cite{Rajagopal:1990yx} is relaxed. In our model 
the axino is not stable. However, if its mass is below the electron mass the only decay 
modes are 
\begin{enumerate}
\item into a neutrino and an axion (tree-level, with dimension 5 operator), $\tilde a \to \nu_i + a$;
\item into three neutrinos (tree-level), $\tilde a \to \nu_i + \nu_j + \nu_k$;
\item into a neutrino and a photon (one-loop), $\tilde a \to \nu_i +\gamma$.
\end{enumerate}
The resulting lifetime, as we show in the following subsections, is longer than the age of
the universe, thus making the axino a dark matter candidate, with relic abundance 
\begin{equation} \label{eq:axinoabundance}
\Omega_{\widetilde a} h^2 \simeq 2.1 \times 10^{-5} \left( \frac{m_{\widetilde a}}{1 \ {\rm keV}} \right) \left( \frac{10^{12} \ {\rm GeV}}{f_a} 
\right)^2 \, .
\end{equation}
This does not depend on $T_{\rm RH}$, as long as $T_{\rm RH} > 10^4$ GeV~\cite{Bae:2011jb, Bae:2011iw}.  
For lower $T_{\rm RH}$ and different cosmological histories the axion abundance can be different~\cite{Co:2015pka}, but 
we will not consider such scenarios in this work.
We learn that, for $f_a = 10^{12}$ GeV, the axino can have a mass as large as a few MeV before it becomes overabundant.
The important feature of Eq.~\eqref{eq:axinoabundance} is that the abundance is inversely proportional to $f_a^2$. It is easy to understand 
why. The axino never reaches thermal equilibrium in the early universe because its interactions are very weak. It is produced in scattering 
processes (listed {\it e.g.} in Table 1 of Ref.~\cite{Bae:2011jb}) with a cross section proportional to $1/f_a^2$. The larger the cross section 
the more the axino is produced, hence the $1/f_a^2$ dependence in Eq.~\eqref{eq:axinoabundance}. In Fig.~\ref{fig:DM} we show the 
allowed region of parameter space, in the plane $f_a$ vs $m_{\widetilde a}$, where we can have both axion and axino dark matter.

\section{Summary} \label{sec:conclusion}
%
We have presented for the first time a complete RpV SUSY model with baryon triality and with a DFSZ 
axion superfield. We 
have studied the mass spectrum and then investigated some cosmological bounds. The axion is a good dark matter 
candidate when its decay constant $f_a = {\cal O} (10^{11})\,$GeV. For lower values of $f_a$ the axino, with a mass 
roughly between 1 and 100~keV, can be the dominant dark matter candidate. We have looked at the possible 
decay modes of the light axino in detail. For such a light axino, its lifetime is longer than the age of the universe, but 
its decay into photon and neutrino still leads to interesting phenomenology. We have shown that X- and gamma-ray 
constraints on this decay give new bounds on some trilinear RpV couplings. These are model dependent and rely on 
the axino constituting the whole dark matter in the universe. We have also shown 
that in this model there is a corner of parameter space where a 7 keV axino could fit the 3.5 keV line observed 
in galaxy clusters.

This model could be explored in further details in relation to neutrino physics or possible collider signatures.
 We leave it to future work.

 \vskip 1cm
\noindent
{\bf Acknowledgements:}  
We thank Branislav Poletanovic for helpful initial discussions, and Toby Opferkuch for comments on various aspects
of this work. We thank Florian Staub for many very helpful 
discussions as well as his assistance in the use of \texttt{SARAH}. LU thanks Francesco D'Eramo for discussions and acknowledges support from the PRIN 
project ``Search for the Fundamental Laws and Constituents'' (2015P5SBHT\textunderscore 002). HKD would 
like to express a special thanks to the Mainz Institute for Theoretical Physics (MITP) for its hospitality and support 
during part of this work.

\begin{appendix}
%
%
%
\section{The Unbearable Lightness of the Axino} \label{sec:lightaxino}
The axino mass has been extensively discussed in the 
literature~\cite{Tamvakis:1982mw, Rajagopal:1990yx, Goto:1991gq, Chun:1992zk, Chun:1995hc}. It is commonly 
held to be highly model dependent, taking on almost any value~\cite{Covi:2009pq, Choi:2013lwa}. In this appendix
we point out that a keV mass axino is more easily embedded in the DFSZ class of models rather than in the KSVZ models~\cite{Kim:1979if, Shifman:1979if}. We then discuss how the mass is 
related to the PQ and SUSY breaking energy scales.

\subsection{KSVZ vs DFSZ}
Axion models fall into two broad categories denoted as KSVZ and DFSZ. Both contain a complex 
scalar field, $A(x)$, charged under a global $U(1)_{\rm PQ}$, PQ symmetry~\cite{Peccei:1977hh}. The axion field,
$a(x)$, is identified with the phase of $A(x)$.\!\footnote{This is not rigorously exact in the DFSZ models, but true to a good 
approximation.} In the KSVZ model, $A(x)$ couples to exotic heavy quark fields, $Q(x)$, that are charged under 
color and under the $U(1)_{\rm PQ}$ (and possibly $U(1)_{\mathrm{EM}}$), while the rest of the SM fields do not 
carry any PQ charge. The coupling $f_Q A \bar Q Q$ generates, via a triangle diagram, the interaction term 
$a G_{\mu\nu} \widetilde G^{\mu\nu}$ which is crucial in solving the strong CP problem. Here $G_{\mu\nu}(x)$ is the gluon field 
strength tensor and $\widetilde G^{\mu\nu}(x)$ its dual. In the supersymmetric version, the KSVZ model is defined by the 
superpotential $W_{\rm KSVZ} = f_Q \hat A \hat{\bar Q} \hat Q$, where we use hats to denote chiral superfields and 
$Q$, $\bar Q$ are two distinct superfields in the \textbf{3} and $\bar{\mathbf{3}}$ representations of $SU(3)_c$, 
respectively, but with equal PQ charge. Embedding the model in supergravity results in a 
one--loop contribution to the axino mass of order~\cite{Moxhay:1984am, Rajagopal:1990yx}
\be
m_{\widetilde a} \sim 10 \ {\rm GeV} \left( \frac{m_{3/2} f^2_Q}{100 \ {\rm GeV}} \right) \, ,
\ee
where $m_{3/2}$ is the gravitino mass. 
Moreover, even if supergravity effects are neglected, there is an irreducible two-loop contribution involving the 
gluino~\cite{Kim:2012bb}
\be
m_{\widetilde a} \sim 0.3 \ {\rm GeV} \left( \frac{m_{\widetilde g} f^2_Q}{1 \ {\rm TeV}} \right) \, ,
\ee
where $m_{\widetilde g}$ is the gluino mass.
This implies that even if the axino mass is tuned to keV values at tree level, loop corrections tend to raise it to GeV values. 
Thus a  KSVZ axino prefers to be heavier than a few keV.

In DFSZ models the field $A(x)$ couples to two Higgs doublets, $H_u$ and $H_d$, which are also charged under the 
$U(1)_{\rm PQ}$, as well as the other SM fields are. Thus the supersymmetric 
version of the DFSZ axion model is more economical than the KSVZ counterpart, as SUSY already requires at least 
two Higgs doublets. The SUSY coupling of axions to Higgses can be written as~\cite{Rajagopal:1990yx} $W_{\rm DFSZ} = 
c_1 \hat A \hat H_u \hat H_d$. The field $A$ has to get a large vacuum expectation value (VEV) $\langle A \rangle \sim f_a$, 
with $f_a > 10^{9}\,$GeV, otherwise the corresponding axion would be excluded by supernova 
constraints~\cite{Raffelt:2006cw}. In turn, this implies a tiny coupling $c_1$. Note that the 
corresponding coupling in non supersymmetric DFSZ models also has to be very small, for the axion to be invisible. In the 
SUSY context it has been proposed that the operator $c_1 \hat A \hat H_u \hat H_d$ could be replaced~\cite{Kim:1983dt} by 
a non-renormalizable one $\frac{g}{M_{\rm Pl}} \hat A^2 \hat H_u \hat H_d$. With the latter operator one easily obtains a 
$\mu$-term at the TeV scale, while with the former we are forced to take very small values the coupling $c_1$. 
We do not address the $\mu$-problem, but we point out that Ref.~\cite{Honecker:2013mya} 
showed that the operator we consider can be derived consistently within a string theory framework. The fact that $c_1$ is 
tiny helps with the axino mass: once we set it to keV values at tree level we are guaranteed that radiative corrections, 
proportional to $c_1$, will be negligible  as opposed to the KSVZ case we discussed above. 

\subsection{Low-scale vs High-scale SUSY Breaking}
In a previous paper~\cite{Dreiner:2014eda}, we indicated a simple way to understand that in models where the SUSY breaking 
scale, $M_{\rm SB}$, is lower than the PQ scale, $f_a$, the axino mass is of order $\mathcal{O} \left( \frac{M^2_{\rm SUSY}}{f_a} 
\right)$, with $M_{\rm SUSY}$ the scale of the soft supersymmetry breaking terms, while in models with $M_{\rm SB} > f_a$ 
the axino mass is typically of order $M_{\rm SUSY}$. The former models, with low $M_{\rm SB}$  are representative of global 
SUSY, for which the best known framework is gauge mediation of supersymmetry breaking \cite{Giudice:1998bp}. The latter 
ones, with high $M_{\rm SB}$ usually fall in the scheme of gravity mediation of supersymmetry breaking, in local 
supersymmetry, and the scale of the soft terms is that of the gravitino mass, $M_{\rm SUSY} \sim m_{3/2}$.

\begin{figure}[!t]
\centering
\includegraphics[width=0.4 \textwidth]{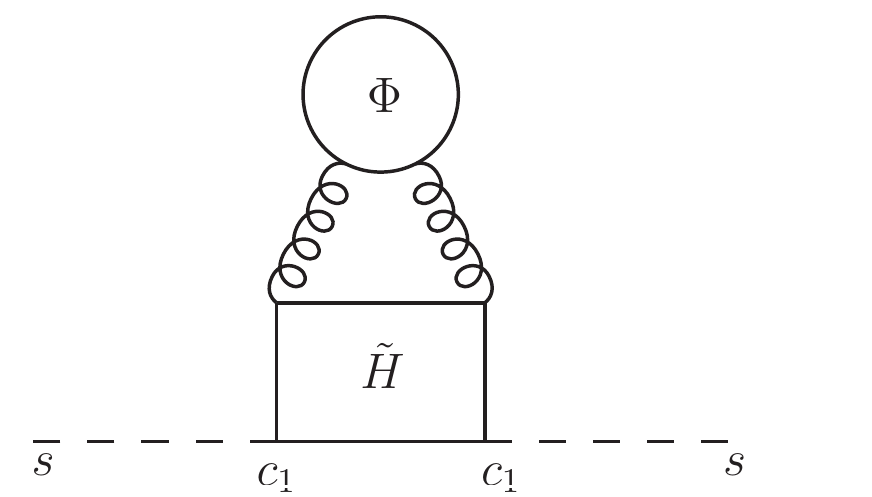}
\caption{Three-loop contribution to the saxion mass in minimal gauge mediation. At the top we have a loop of messenger fields, 
$\Phi$, drawn as a circle, at the bottom a loop of Higgsinos, drawn as a rectangle. The wiggly lines are gauge bosons. The 
coupling $c_1$ of the saxion to the Higgsinos is defined in Eq.~\eqref{eq:WI}.}
\label{fig:3loops}
\end{figure}

In light of these considerations one would think that a light axino is more natural in the context of gauge mediation. However it turns 
out that this is difficult to accommodate. The problem is with the saxion. Consider a model of minimal gauge mediation (see {\it 
e.g.} Ref.~\cite{Giudice:1998bp}), where the messengers do not carry any PQ charge and communicate with the visible sector only 
via gauge interactions. The leading contribution to the mass of the sfermions of the minimal supersymmetric Standard 
Model (MSSM) is generated at two loops. However the saxion is a gauge singlet and its mass can only be generated at three
loops, as shown schematically in Fig.~\ref{fig:3loops}.
The coupling $c_1$ that appears twice in the diagram is the one we discussed in the previous subsection, and the same that appears 
in the superpotential of Eq.~\eqref{eq:superpotential}. The saxion squared mass, $m_s^2$ is suppressed by a factor of $\sim \frac{c_1^2}
{16 \pi^2}$ compared to the squared masses of the MSSM sfermions, $m_{\widetilde f}^2$, so we can estimate
\be
m_s \sim 0.1 \ {\rm keV} \ \left( \frac{c_1}{10^{-9}} \right) \left( \frac{m_{\widetilde f}}{1 \ {\rm TeV}} \right) \, .
\ee 
This light saxion could pose serious cosmological problems~\cite{Banks:2002sd} as it would come to dominate the energy density 
of the universe for a long time before it decays. We reviewed some of the issues associated with saxion cosmology in  
Sec.~\ref{sec:cosmology}. One way out of this problem would be to make the saxion heavier. This could be achieved by either coupling the axion 
superfield to the messengers or to fields in the hidden sector responsible for SUSY breaking. Apart from the extra field content that 
the procedure would bring into the model there is another issue that seems difficult to overcome: the same couplings needed to 
make the saxion heavier would very likely produce a heavy axino~\cite{Carpenter:2009sw}. Perhaps there is a clever way to 
arrange for a heavy saxion and a light axino in gauge mediation, but in light of our considerations it seems that such a model would 
have to be quite complicated. We do not investigate this aspect further in this work. Rather we choose a supergravity (SUGRA) 
model. We showed in Ref.~\cite{Dreiner:2014eda} that in SUGRA the axino would typically have a mass comparable to the 
gravitino, thus in the TeV range. However, we can adjust a single parameter, $\lambda_{\chi}$, to lower the axino 
mass down to the keV range. Doing so does not affect other masses, such as those of saxion and gravitino for instance. Also, as 
we argued above, since the axino we consider is of the DFSZ type, once we set its mass at tree level it will not be affected 
appreciably by loop corrections. Therefore the SUGRA model can be kept minimal, as opposed to a possible model with gauge 
mediation, at the expense of some tuning, needed to lower the mass of the axino. In the model presented in this paper the 
axino is much lighter than the gravitino. This represents an explicit exception to the generic argument that the axino should be 
heavier than the gravitino, put forward in Ref.~\cite{Cheung:2011mg}. 

\section{Axino-Gaugino Mixing} \label{sec:axino-mixing}
%
\begin{figure}[t] 
\includegraphics[width=\textwidth]{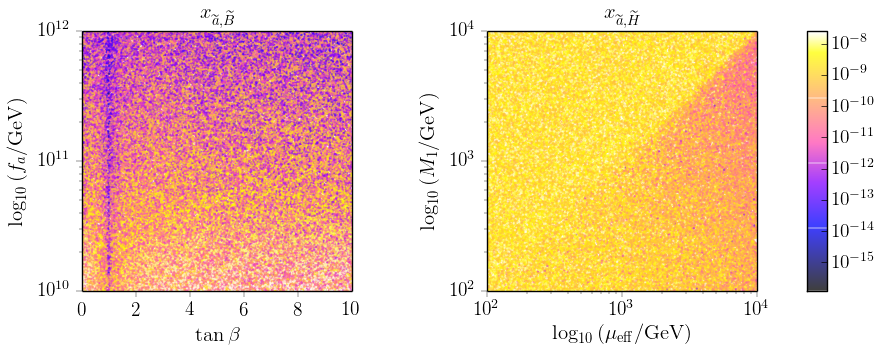}
\caption{Distribution of the mixings $x_{\widetilde{a},\widetilde{B}}$ and $x_{\widetilde{a},\widetilde{H}}$ 
obtained by numerical evaluation of the eigenvalues of $\mathcal{M}_4$. We run $10^5$ points randomly varying 
$10^2\, {\rm GeV}\, < M_{1\,,2},\mu_{\rm eff}<10^4 \,{\rm GeV}\,,\,1<\tan\beta<10\,,\,10^{10}\,{\rm GeV}<f_a<10^
{12}\,{\rm GeV}\,,\,1\,{\rm keV}<m_{\widetilde{a}}<10^2\,{\rm keV}$. From our estimate in Eq.~\eqref{ahimix} 
[Eq.~\eqref{abimix}] we expect $\mixH$ [$\mixB$] to lie somewhere in the range between approximately $10^
{-10}$ and $10^{-8}$ GeV [$10^{-12}$ and $10^{-10}$ GeV], depending on the value of $f_a$.} \label{fig:AxinoGauginoMixings}
\end{figure}
Most of the amplitudes involved in the radiative decay of the axino bear a strong dependence on its mixing with some 
of the other neutralino mass eigenstates. In this work we rely on well-motivated estimates for these quantities, but in 
order to obtain their precise values one should diagonalize the neutralino mass matrix $\mathcal{M}_{\chi^0}$. While 
an exact analytical algorithm exists for the case of the MSSM~\cite{Choi:2001ww}, adding one more mass eigenstate, 
as for the case of the Next-to-Minimal Supersymmetric Standard Model (NMSSM), does not allow for a closed-form 
solution. In this section we show how the approximate block-diagonalization procedure of Ref.~\cite{Choi:2004zx} for the 
NMSSM 5x5 neutralino mass matrix can be used to obtain a numerical value for the axino-gaugino mixing. The only 
requirement is a small mixing between the the singlet and the doublet states,  which in our case corresponds to the 
mixing between any of the two Higgs doublets and the axion field $A$. From $\mathcal{M}_{\chi^0}$ in 
Eq.~(\ref{eq:Mchi10}) we see that such mixing is indeed small, since $\frac{c_1 v_{u,d}}{\sqrt{2}} \approx \frac{\mu_
{\rm eff}}{f_a} \lesssim 10^{-7}$. Recalling that the axino mass eigenstate consists of a linear combination of the 
fermionic components of the fields $A$ and $\bar{A}$ only, we define the following neutralino mass 5x5 sub-matrix:
\begin{align}
\mathcal{M}_5&=\left[ 
\begin{array}{ccccc}
M_1 &0 &\frac{g_1 v_u}{2}   &-\frac{g_1 v_d}{2} &0\\
0 &M_2 &-\frac{g_2 v_u}{2}   &\frac{g_2 v_d}{2} &0\\
\frac{g_1 v_u}{2}   &-\frac{g_2 v_u}{2}  &0 &- \frac{c_1 v_A}{\sqrt{2}} &- \frac{c_1 v_d}{\sqrt{2}} \\
-\frac{g_1 v_d}{2}   &\frac{g_2 v_d}{2}   &- \frac{c_1 v_A}{\sqrt{2}}   &0 &- \frac{c_1 v_u}{\sqrt{2}} \\
0& 0&- \frac{c_1 v_d}{\sqrt{2}}   &- \frac{c_1 v_u}{\sqrt{2}} &m_{\widetilde{a}} \\
\end{array} \right]\,,
\end{align}    
where the axino mass $\maxino$ is the lightest eigenvalue which results from the diagonalization of the 3x3 
lower-right block of the neutralino mass matrix $\mathcal{M}_{\chi^0}$ in Eq.~\eqref{eq:Mchi10}. The diagonalization of 
$\mathcal{M}_5$ proceeds in two steps: first, the 4x4 matrix $V$ rotates only the MSSM upper-left 4x4 block into a diagonal 
form; next a second matrix block-diagonalizes the full resulting 5x5 matrix, in a  way that the 4x4 MSSM sub-matrix 
$\mathcal{M}_4$ is still diagonal up to corrections of the second order in the singlet-doublet mixing parameter $\frac{\mu
_{\rm eff}}{f_a}$. The 5x5 unitary matrix $U$ which combines these two steps is 
\begin{align}
U &= \left[
\begin{array}{cccc}
\mathbb{1}_4-\frac{1}{2}\left(V \Lambda\right)\left(V \Lambda\right)^T  &\left(V \Lambda\right) \\
-\left(V \Lambda\right)^T &1-\frac{1}{2}\left(V \Lambda\right)^T \left(V \Lambda\right) \\
\end{array}
\right] \left[ 
\begin{array}{cc}
V &0 \\
0 &1 \\
\end{array}
\right] \,,
\end{align} 
with
\begin{align}
\Lambda=-\frac{m_{\tilde{a}}}{\det\left({\mathcal{M}_4-\mathbb{1}_4 m_{\widetilde{a}}}\right)}\left[
\begin{array}{c}
\left(M_2-m_{\widetilde{a}}\right) g_1\frac{v}{2} \mu_{\rm eff}\cos 2 \beta \\
\left(M_1-m_{\widetilde{a}}\right) g_2\frac{v}{2} \mu_{\rm eff}\cos 2 \beta \\
\left(M_1-m_{\widetilde{a}}\right)\left(M_2-m_{\widetilde{a}}\right)\left(\mu_{\rm eff} \sin\beta-m_{\widetilde{a}}\cos\beta\right)-M_{\star}^{3}\cos\beta\\
\left(M_1-m_{\widetilde{a}}\right)\left(M_2-m_{\widetilde{a}}\right)\left(\mu_{\rm eff} \cos\beta-m_{\widetilde{a}}\sin\beta\right)-M_{\star}^{3}\sin\beta\\
\end{array}
\right]\,,
\end{align}
 and $M_{\star}^{3}=\frac{v^2}{4}\left[
\left(M_1-m_{\widetilde{a}}\right)g_2^2+\left(M_2-m_{\widetilde{a}}\right)g_1^2\right]$. 
The axino-gaugino mixings can then be read off from the off-diagonal entry $\sum_{j=1}^4 V_{ij}\Lambda_j$ of $U$, 
with $j=\widetilde{B}_0,\widetilde{W}_0,\widetilde{H}_u^0,\widetilde{H}_d^0$. In Fig.~\ref{fig:AxinoGauginoMixings} we 
show two examples of how a more careful analysis can change significantly such quantities, and thus ultimately the 
phenomenology of the model. This possibly constitutes caveats to our above reasoning. In the left panel we observe that a 
value of $\tan\beta$ close  to 1 strongly suppresses the value of $x_{\widetilde{a},\widetilde{B}}$, such that 
our previous estimates are off by several orders of magnitude. In the right panel instead we see how allowing for larger 
values of $\mu_{\rm eff}$ might lower $x_{\widetilde{a},\widetilde{H}}$, if $M_1 < \mu$.    
\end{appendix}

\bibliographystyle{JHEP}
\bibliography{Axino}

\end{document}